\documentclass[12pt]{article}

\usepackage{amssymb}
\usepackage{amsmath}
\usepackage{epsfig}

\def\be{\begin{equation}}
\def\ee{\end{equation}}
\def\bea{\begin{eqnarray}}
\def\eea{\end{eqnarray}}

\begin{document}
\begin{titlepage}
\thispagestyle{empty}
\hskip 1 cm
\vskip 0.5cm

\vspace{25pt}
\begin{center}
    { \LARGE{\bf Archipelagian Cosmology}}

 \vspace{15pt} 

{\LARGE{or}}

\vspace{15pt}

{ \LARGE{\bf Dynamics and Observables in a
    Universe with Discretized Matter Content}}
    \vspace{33pt}

  {\large  {\bf   Timothy Clifton\footnote{tclifton@astro.ox.ac.uk} and Pedro
  G. Ferreira\footnote{p.ferreira1@physics.ox.ac.uk}}}

    \vspace{15pt}

 {Department of Astrophysics, University of Oxford, UK.}

\vspace{15pt}

{5th of November, 2009}

\vspace{15pt}

  \end{center}

   \vspace{20pt}

\begin{abstract}

We consider a model of the Universe in which the matter content is in
the form of discrete islands, rather than a continuous fluid.  In the
appropriate limits the resulting large-scale dynamics approach those of a
Friedmann-Robertson-Walker (FRW) universe.  The optical
properties of such a space-time, however, do not.  This illustrates the fact that
the optical and `average' dynamical properties of a relativistic
universe are not equivalent, and do not specify each other uniquely.
We find the angular diameter distance, luminosity distance and
redshifts that would be measured by observers in these space-times, using both
analytic approximations and numerical simulations.  While different from
their counterparts in FRW, the effects found do not look like promising
candidates to explain the observations usually attributed to the
existence of Dark Energy.  This incongruity with standard FRW cosmology
is not due to the existence of any unexpectedly large structures or
voids in the Universe, but only to the fact that the matter content of
the Universe is not a continuous fluid.

\end{abstract}

\vspace{10pt}

% \end{titlepage}

\tableofcontents
\thispagestyle{empty}

\end{titlepage}

% \newpage

%\psfrag{3}{$3$}
%\begin{figure}[ht]
%\center \epsfig{file=*.eps,height=6.5cm} 
%\caption{{\protect{\textit{}}}}
%\label{}
%\end{figure}

\section{Introduction}

One of the most basic assumptions made in modern cosmology is the idea that
the geometry of space is homogeneous and isotropic, resulting in the line-element
\be
\label{FRW}
ds^2=-dt^2+a^2(t) \left( \frac{dr}{1-k r^2} + r^2
(d\theta^2+\sin^2\theta d\phi^2)\right),
\ee
where $k$ is the conformal curvature of
space, and $a(t)$ is the time-dependent scale factor of the Universe.
The simplification that results from this assumption is
remarkable: Einstein's equations, a set of 10 coupled, non-linear
partial differential equations in 4 variables, reduce to a single
ordinary differential equation in one variable:
\be
\label{Fried}
\frac{\dot{a}^2}{a^2} = \frac{8 \pi \rho}{3}-\frac{k}{a^2}.
\ee
The over-dot here is a derivative with respect to time, and $\rho$ is
the (spatially constant) energy density of the continuous fluid that
is assumed to permeate the whole of space.  Such models are known as
Friedmann-Robertson-Walker (FRW) cosmologies, and are ubiquitous.

The unreasonable simplicity offered by the assumptions outlined above
has allowed enormous progress to be made in understanding the cosmological models
that result from them, and the ways in which they can be compared to
astronomical observations.  So great is this success that there now
exists a growing sense that such models must be the {\it only} ones
that are suitable to describe our Universe.  These sentiments are
bolstered by observations of the Cosmic Microwave Background (CMB),
which is isotropic to within one part in a hundred thousand.  

However, albeit effective, the homogeneity and isotropy of space is
still only an assumption, and the isotropy of the CMB, while
consistent with a homogeneous and isotropic space-time, does not necessitate
it \cite{EllisCop}.  Furthermore, it now seems to be the case that these assumptions lead inevitably, and
unenviably, to the conclusion that the
Universe should be filled with an exotic fluid that behaves 
repulsively under gravity. This fluid, dubbed {\it Dark Energy}, has
to make up $\sim 2/3$ of the total energy budget
of the Universe and should be responsible for driving a 
late period of apparently accelerating expansion \cite{Sne1,Sne2,BAO,WMAP}. Unfortunately,
the existence of such a fluid provides colossal theoretical challenges,
and despite ongoing efforts to find it, it has yet to be detected
directly. It therefore seems natural to retrace our
steps, and to scrutinise more critically the assumptions that led us
to deduce its existence in the first place.

Inhomogeneous cosmology has been an active field of research for many
decades now \cite{krasinski}. One of the simplest
examples, which places us directly at the centre of the Universe, is the Lemaitre-Tolman-Bondi (LTB) model: A
spherically symmetric, and dust dominated model. Once again it
is implicitly assumed that there exists a continuous fluid that
permeates all of space-time. It is also usual to assume that the spherical
inhomogeneity in this model takes the form of a giant under-density at the
centre of the Universe - i.e. that we live in a void. 
LTB models have been extensively studied over the past few years, and
their behaviour and optical properties are now becoming well understood \cite{voids1}-\cite{voids19}. They can mimic
an FRW universe with Dark Energy, but are required to have a fairly
intricate structure in order to be able to simultaneously match
supernova observations and measurements of primary anisotropies in the CMB. 

A natural next step is to consider multiple LTB or Schwarzschild
space-time patches, embedded in an FRW background.  The
inhomogeneities that result are then part of a continuous, and otherwise constant
fluid that constitutes an exact solution of Einstein's equations.  Such a
Universe could conceivably be treated as being homogeneous and
isotropic on the largest scales, but is endowed with substantial
inhomogeneity on smaller scales. Dubbed as ``Swiss Cheese'', these
models have been explored in great detail; from the ground breaking
work of Kantowski \cite{Kant}, up to the present day \cite{Swiss1}-\cite{Swiss5}.  Again, substantial effort
has gone into understanding the optical properties of Swiss Cheese
universes, and they have been shown to be broadly similar, but not identical, to those of FRW.

Beyond the realm that can be easily investigated with exact solutions,
one will also be interested in approximations that may allow further
insight to be had.  One scheme for including inhomogeneity without
the aid of exact solutions was provided by Dyer and Roeder
\cite{DR1,DR2}.  They ignored the influence of shear on the evolution of
bundles of null geodesics, and taking FRW values for the Hubble rate
and redshift found a neat method of approximating observables in an
inhomogeneous universe.  This method was generalised further by
Mattson \cite{Matt}, who included the effect of an inhomogeneous
Hubble rate.  It was claimed in \cite{Matt} that with an appropriately
changing $H(z)$ the effect of inhomogeneities could entirely account for the deviations from
Einstein-de Sitter (EdS) cosmology that are usually attributed to Dark Energy.  However, a lack of
any model on which to base these results, and the {\it ad hoc} way in
which inhomogeneous expansion is dealt with, makes these claims appear
somewhat speculative.

A further avenue of research that has attracted considerable interest
is the idea that the structures that form in the Universe could have a
`back-reaction' effect on the cosmological expansion \cite{back1}-\cite{back17}.  The
idea here is based on the fact that the operation of averaging the geometry
of space does not, in general, commute with the operation of evolving a space-like
surface forward in time.  Such complications mean that when evolving forward
an `averaged' homogeneous and isotropic space one should include
corrections to Einstein's equations (which are valid for the true
geometry, and not the averaged one).  Some have argued that the
results of these corrections may be large enough to entirely account for Dark
Energy, while others maintain the opposite position.  These studies
are considerably complicated by the implicit difficulties involved with
averaging the geometry of space in General Relativity.

In this paper we choose to take a different path, and completely break with FRW space-time.
We want to work with a model of the Universe in which all of the matter content is in the form
of discrete islands of mass, scattered in otherwise empty space. 
To a first approximation, this appears to be what we see on the night sky: Scattered points
of light, organised into a variety of structures and
meta-structures. Our Universe, at its most basic, is made up of
galaxies with a typical mass roughly that of the Milky Way, dominated
by dark matter and with a number density of $\sim 1$ galaxy per cubic
mega-parsec.

Such a universe should have distinctive properties. Light will propagate in
the empty spaces between the islands, and will no longer pass through
a continuous fluid.  The geometry of these two situations is very
different, and should be expected to result in different optical
properties, even if the large-scale dynamics are equivalent \cite{Zeld}.
To see why this is so, consider the form of the Sachs optical
equations \cite{Sachs} (which we will discuss in detail later on):
\bea
\frac{d \tilde{\theta}}{d \lambda} +\tilde{\theta}^2+ \sigma^2 &\sim& R\nonumber \\
\frac{d \sigma}{d \lambda} +2 \sigma \tilde{\theta} &\sim& C \nonumber.
\eea
Here $\tilde{\theta}$ and $\sigma$ are the expansion and shear scalars, respectively, $C$
represents the Weyl tensor, and $R$ the Ricci tensor. The angular diameter and luminosity
distances are then given by integrals of $\tilde{\theta}$. In a universe where light travels primarily through empty
space, the driving terms in the equations above will be $R=0$ and
$C\neq0$.  The corresponding terms in a spatially flat FRW universe will be $C=0$,
due to its conformal equivalence to Minkowski space, and $R\neq0$, due to its continuous
mass distribution.  These two fundamentally different types of
curvature can be seen to have correspondingly different effects on
bundles of null geodesics, and hence these effects should be taken into account in
observational cosmology.

It is, or course, a very difficult proposition to calculate the
geometry of space-time associated with arbitrarily scattered islands
of matter.  Fortunately, if we return to the underlying principle behind FRW (that we do not live in a special place in
the Universe), and apply it to a discretized matter distribution, then
there is a way forward. In a seminal paper,
Lindquist and Wheeler \cite{LW} showed that it is possible to construct an approximation to a cosmological
space-time by considering a regular lattice of Schwarzschild
space-times with metrics given by the line-element
\be
\label{Schw}
ds^2=-\left(1-\frac{2 m}{r} \right) dt^2 + \frac{dr^2}{\left( 1 -
 \frac{2 m}{r} \right)} + r^2 (d\theta^2+\sin^2\theta d\phi^2).
\ee 
They did so for a closed universe and found that, in the limit of
 a large number of masses, such a space-time evolves almost identically
 to a closed FRW Universe: The separation between islands of mass
 increases, reaches a maximum, and then contracts with exactly the
 same time dependence as one would find for a perfectly smooth, dust
 filled universe.  Configurations of gravitating discrete masses have
 also been considered in \cite{battye}.

In this paper, we wish to explore the Lindquist-Wheeler model further,
and to investigate the optical properties of such a universe. It is
not the goal of this paper to propose a complete alternative
to the currently favoured FRW universe dominated by a cosmological
constant. Hence, we will restrict ourselves to considering only regular
distributions of masses, and often only to the equivalent of an EdS
universe\footnote{Even though this assumes a matter
content that is almost three times that which has apparently been observed through dynamical estimates
with clusters and peculiar velocities, using FRW relations between
redshift and distance.}. We expect that such an analysis
should allow us to uncover some of the essential features of these
space-times, even if it is not the most general case possible.  We
will return to the issue of extending our study to other spatial
curvatures, matter contents and distributions of mass in future publications.

The paper will proceed as follows:

In Section \ref{approx} we recap the lattice model proposed by Lindquist
and Wheeler in 1957 \cite{LW}.  
%We discuss their motivation for considering such a model 
%which was based on the success of the Wigner-Seitz
%construction of electro-magnetism.  We then reproduce some known
%results stating which lattice structures can be formed from regular
%polyhedra in 3-spaces of constant curvature, before proceeding to
We reproduce their main result that the dynamics of a lattice
in a closed 3-space obeys an evolution equation with the same
functional form as the Friedmann equation, and with a %different
scale that approaches the FRW one as the number of points in the lattice is increased. %this
%scale approaches the corresponding FRW value, and the dynamics become
%identical to FRW.  
We then extend Lindquist and Wheeler's study to
include more general space-times, with arbitrary spatial curvature, and with a cosmological
constant.  We extend the `cosmological time' coordinate they developed
away from just the boundaries of the cells, giving us a global coordinate
system with which to study the propagation of photons in, and
between, cells.  The problem of over-lapping Schwarzschild space at
a boundary is discussed.  The dynamics of the lattice is then shown to
behave analogously to flat, closed, and open FRW universes, with or without a
cosmological constant, when the curvature of the 3-space in which the
lattice is constructed is chosen appropriately.

With our approximate space-time in hand, we then proceed in Section
\ref{3} to investigate null trajectories within it.  The relevant null
geodesic equations are presented in the coordinate system found in
Section \ref{approx}.  We explain how to propagate light rays between
lattice cells using two different methods; one simple and approximate,
the other more elaborate, and accurate.  It is found that the results
we present in later sections are reasonably insensitive to which particular
method is chosen.

We are then in a position to study redshift, as measured by
observers looking along the null trajectories discussed above.  In Section
\ref{4} we present the formalism needed to determine these shifts, and
make both analytic and numerical approximations to their solutions.
The measured redshift to a radiating source is found, in general, to approach a
value that is smaller than the corresponding shift in FRW, with
relatively little scatter around the mean.  The relation satisfied is $1+z
\simeq (1+z_{FRW})^{7/10}$.  It is also found that deviations
from this rule occur for trajectories that follow a special direction (such as being confined to a
plane that picks out a symmetry of the lattice).

In Section \ref{5} we solve the Sachs optical equations along our null
geodesics.  These equations determine the expansion and shear of a
bundle of null rays that are focused at either the source or
observer.  We provide a method of propagating these quantities between
cells, and make analytic and numerical approximations of the distance
measures that result.  We also note the importance of the caustics that can develop due to the
influence of shear.

Section \ref{6} contains a discussion of the observational
consequences of the preceding sections.  In terms of redshift, we
find that the luminosity distance for a spatially flat dust dominated universe with
$\Lambda=0$ takes the form $r_L \propto (1+z)^2-(1+z)^{-\frac{8}{7}}$, which corresponds to a deceleration parameter of
$q_0=8/7$ (i.e. objects at the same $z$ appear {\it brighter} than in
EdS).  In terms of cosmological time, however, objects at the same $t$
can appear dimmer.  We present the Hubble diagram for this space-time,
in the form of a plot of distance modulus, and continue to speculate
on other cosmological observables such as CMB anisotropies, baryon
acoustic oscillations and galaxy number counts.

In Section \ref{7} we conclude. A series of appendices then follow.

Throughout the paper we attempt to present as many of our results as
possible using analytic methods. We back this up with Monte Carlo
simulations from a ray tracing code that allows us to confirm their
accuracy. 

\section{An Approximate Space-Time}
\label{approx}

Inspired by the success of the Wigner-Seitz construction in
electro-magnetism \cite{WS1,WS2}, Lindquist and Wheeler (LW) constructed in
\cite{LW} a lattice model of the Universe.  The ideas they put forward
are of central importance to our study, and so we reiterate them here.

\subsection{The Lindquist-Wheeler Model}
\label{1}

Starting with a positively curved hyper-sphere, LW distributed a number of
``mass concentrations'' into a regular lattice that they formed from
tiling the 3-space with regular polyhedra\footnote{Tilings of 3-spaces
of constant curvature are discussed in Appendix A.}.  Each cell of the lattice was given
a central mass, and then approximated by a sphere, with the true
geometry of the space-time (that which would
result from solving Einstein's equations) being replaced by the
Schwarzschild geometry of the closest mass.  As LW noted: ``This
approximation demands that the distribution of gravitational
influences just external to each sphere should depart relatively
little from spherical symmetry''.  

The accuracy of this approach can
be evaluated in the Wigner-Seitz construction by comparing to known
exact solutions \cite{WSexact1,WSexact2}, with favourable results.
Such solutions involve ``empty lattices'', where the potential is
taken to be a constant throughout. In General Relativity a similar
test can be performed, but this time with a cosmological constant
dominating the gravitational interaction \cite{next}.  The lattice results can again be seen to
approach the exact solution (de Sitter space) in the appropriate
limits, lending credance to the lattice model.

The essential difference between the gravitational and electro-magnetic
cases, apart from the non-linearity of the field equations in
General Relativity, is that in the lattice model constructed by LW,
the lattice itself is dynamical.  This is because the non-zero normal
derivative of the gravitational potential at
the cell boundary results in a relative motion between the
boundary and central mass.  It is this motion that LW concerned themselves with,
and with which they constructed an approximate global, dynamical
space-time out of the Schwarzschild solution alone.

%As discussed above, the non-zero normal derivative of the
%gravitational potential at the cell boundaries results in a dynamical
%lattice.  
For a cosmological interpretation, the scale factor of FRW cosmology
now has to be replaced by some measure of the `size' of the lattice, and the Hubble rate has to be
replaced by the rate of increase in this size.  To go further we must
therefore explain what measure of size is intended, and with
which time coordinate the expansion rate is defined with respect
to. A key concept in the LW model, in this respect, is the idea of
tangency between the constant curvature background hyper-sphere (on
which the lattice is defined), and the 3-spaces that are locally
orthogonal to the trajectory of the boundary sphere of each cell.
This tangency gives a natural time coordinate, $\tau_{LW}$, with which to describe the expansion (at least in
the vicinity of the bounding spheres).  What is more, it is a time
coordinate that has some global meaning, as it can be used to define a congruence of
time-like trajectories that are orthogonal to a common 3-space.  As LW build their lattices in
hyper-spherical 3-spaces, they also have a
natural measure for their size:  The radius of the hyper-sphere in an
embedding Euclidean 4-space, $a_{LW}$.

It was found by LW that, in terms of the coordinates discussed above, the
dynamics of their lattice is specified by an evolution equation of
the form
\be
\frac{\dot{a}^2_{LW}}{a_{LW}^2} = \frac{2 m}{a_{LW}^3  \sin^3 \psi} - \frac{1}{a_{LW}^2},
\ee
where the over-dot represents a derivative with respect to
$\tau_{LW}$, $m$ is the Schwarzschild mass at the centre of each cell, 
and $\psi$ is the (constant) angle subtended at the centre
of the hyper-sphere between vectors in the Euclidean embedding 4-space
that connect the centre of the hyper-sphere with the centre and
spherical boundary of one cell.  Clearly, this equation has the same
functional form as the Friedmann equation, (\ref{Fried}), with a
dust-like energy content, and positive spatial curvature.  

%\begin{table}
%\begin{center}
%\begin{tabular}{|c|c|c|}
%\hline
%\bf{N} & $\mathbf{\sin \psi}$ & $\mathbf{
%  a_{LW}^{(max)}/a_{FRW}^{(max)}}$ \\ 
%\hline
%5 & 0.871 & 1.43\\
%8 & 0.773 & 1.28 \\
%16 & 0.634 & 1.16\\
%24 & 0.561 & 1.12\\
%120 & 0.336 & 1.04\\
%600 & 0.198 & 1.01 \\
%$\infty$ & 0 & 1.00 \\
%\hline
%\end{tabular}
%\end{center}
%\caption{{\protect{\textit{The number of cells in each lattice, $N$, the angle $\psi$
%  subtended by each cell at the origin of the hyper-sphere, and the
%  maximum of expansion as a fraction of the FRW value.  The last row
%  indicates the FRW limit.}}}}
%\label{table2}
%\end{table}

\begin{figure}[t]
\center \epsfig{file=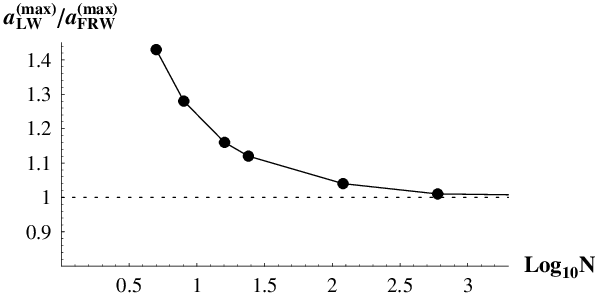,height=6cm} 
\caption{{\protect{\textit{The maximum radius of expansion, as a
	fraction of the FRW value, for lattices with $N=5,8,16,24,120$
	and $600$ cells.  The dashed line corresponds to the maximum
	of expansion in a spatially closed, dust dominated FRW universe.}}}}
\label{LWplot}
\end{figure}

The LW lattice therefore evolves in a similar way to a dust dominated, closed
FRW universe.  The only difference is the value of the maximum radius
of the universe, when expansion ends and the onset of collapse is
about to begin. In the LW case the maximum radius is given by
\be
a_{LW}^{(max)} = \frac{2 M}{N \sin^3 \psi},
\ee
where $M=N m$ is the total mass of all cells in the lattice.  In the
FRW case we have
\be
a_{FRW}^{(max)} = \frac{4 M}{3 \pi},
\ee
where the total mass $M$ is now written in terms of the energy
density from the Friedmann equation as\footnote{The surface
area of a hyper-sphere of radius $r$ is $2 \pi^2 r^3$.} $M=2 \pi^2 a_{FRW}^3 \rho$.  The ratio of
these two maximum radii is clearly independent of the total mass, $M$,
in the space-time, and depends only on the number of cells in the
lattice, $N$, and the angle they subtend at the centre of the
hyper-sphere, $\psi$.  It provides a measure of the departure from
FRW evolution, with the dynamics of the lattice approaching FRW in the
limit $a^{(max)}_{LW}/a^{(max)}_{FRW} \rightarrow 1$.

We know all possible values of $N$ from Appendix A, and can
straightforwardly work out the value of $\psi$ for each lattice once
it has been specified how the lattice cell polyhedra are to be
replaced by spheres.  LW considered two possible generalizations of
the Wigner-Seitz approximation to curved space: (I) That the boundary
sphere of each cell should occupy the same volume as the cell, and
(II) that the bounding sphere should be just large enough to touch its
nearest neighbours.  The results of (I) are shown 
graphically in Figure \ref{LWplot}.  It
can be seen that as $N \rightarrow \infty$, and the continuum limit is
approached, the lattice approaches the dynamical evolution of a
spatially closed, and dust dominated universe.  At $N=600$ the
difference is already less than $1.5\%$.  This is the main result of LW.

%\subsubsection{Lattice dynamics}

%\section{An Approximate Space-Time}
%\label{approx}
%
%LW considered the dynamics of a lattice universe in a closed space,
%and found it to be similar to the corresponding perfect fluid model.
%We now wish to generalise their model to include other spatial curvatures,
%and, crucially, to allow us to calculate the optical properties of the space-time.

\subsection{Geometric set-up}

The LW construction was engineered to allow the geometry inside
individual cells to be approximated by the Schwarzschild
solution, (\ref{Schw}).  Gluing these cells together in a suitable way
then gives a global, dynamical lattice space-time.
They found the dynamics of a lattice universe in a closed space,
and showed them to be similar to the corresponding perfect fluid model.
We now wish to generalise their model to include other spatial curvatures,
and, crucially, to allow us to calculate the optical properties of the space-time.

\begin{figure}[t]
\center \epsfig{file=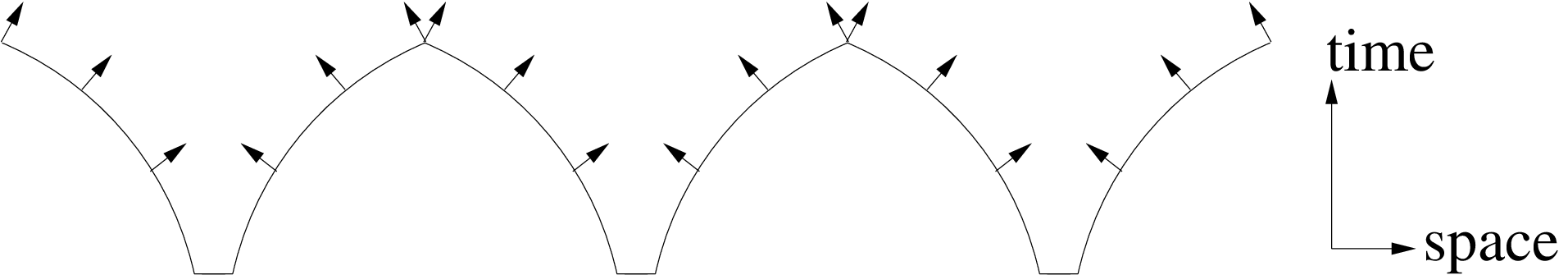,height=2.5cm} 
\caption{{\protect{\textit{A schematic of three one-dimensional space-like
	regions with intersecting, but not overlapping, surfaces of
	constant time, placed next to each other.  At the boundary
	between cells different normal time-like vectors from different cells
	point in different directions.  This is the way surfaces of
	constant $t$ behave in our model. Hence, $t$ does not represent
	a good choice for a global, cosmological time coordinate.}}}}
\label{cells}
\end{figure}

The Schwarzschild coordinates are, of course, perfectly acceptable when
considering the geometry inside a single cell.  They are not, however,
well suited to describing the global geometry of a universe with many
discrete sources.  The problem is that these coordinates will not mesh
at cell boundaries - that is, the 3-space of constant $t$ from one
cell will not be tangent to the  corresponding 3-space from
any other.  This makes gluing cells together problematic, as the space
that we would construct by putting cells at constant $t$
next to each other would not be at all smooth:  The coordinate
patches for the spaces would intersect, rather than overlap, making
any global interpretation of `time' very difficult indeed.  This
problem is illustrated in Figure \ref{cells}, with 3 intersecting
1-dimensional spaces placed next to each other.

\begin{figure}[t]
\center \epsfig{file=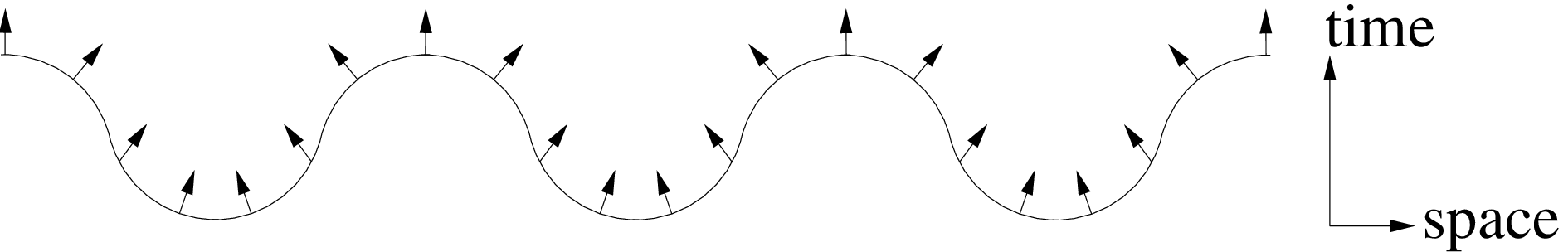,height=2.3cm} 
\caption{{\protect{\textit{An example of three space-like regions with
overlapping, rather than intersecting, surfaces of constant time.
The normal time-like vectors from different cells can now be
identified at the boundary.  This is the way surfaces of constant
$\tau$ behave, and hence $\tau$ can be interpreted globally.  The smooth
cell bottoms here represent the non-divergent behaviour of the
$\tau=$constant surfaces at $r=2 m$.}}}}
\label{cells2}
\end{figure}

The problem was solved in LW by introducing a new time coordinate in the
vicinity of the spherical cell boundaries that ensured the 3-spaces of
constant time were orthogonal to the trajectory of the boundary.  This allowed the two space-like surfaces
of two adjoining cells to be tangent at any point where the boundaries
met, hence allowing a common definition of time.  This situation is
illustrated in Figure \ref{cells2}.
We now want to generalise their coordinates to spaces
with different spatial curvatures, and away from the boundaries so that we can
propagate null trajectories all the way through each cell in a
consistent way.

To do this, first consider a single Schwarzschild cell with a
spherical boundary in free-fall.  Now perform the following
transformation from Schwarzschild time, $t$, to a new time coordinate, $\tau$:
\be
\label{tau}
d \tau = \sqrt{E} dt- \frac{\sqrt{E-\left(1- \frac{2 m}{r}
 \right)}}{\left( 1- \frac{2 m}{r}\right)} dr,
\ee
where $E$ is a positive constant.  The Schwarzschild line-element
 (\ref{Schw}), describing the geometry inside the cell then becomes
\be
\label{pan}
ds^2 = -\frac{1}{E}\left(1- \frac{2 m}{r} \right) d \tau^2 - \frac{2}{E}
\sqrt{E-\left(1-\frac{2 m}{r} \right)}
d\tau dr +\frac{dr^2}{E} +r^2 d\Omega^2.
\ee
%where $d\hat{s} = \sqrt{E} ds$.
In the limit $E\rightarrow 1$ this reduces to the well known Gullstrand-Painlev\'{e}
coordinates \cite{pan1,pan2}.  The trajectory of a radially out-falling time-like
geodesic is then given by
\be
\label{tl}
\left( \frac{dr}{d\tau} \right)^2 = (E-1)+\frac{2 m}{r},
\ee
where $\tau$ is also the proper time measured along the trajectory\footnote{For
in-falling trajectories one should take the opposite sign for the
square roots in (\ref{tau}) and (\ref{pan}).}.  
%Normalising with respect to the geometry specified by $d\hat{s}^2$, e
Each free-falling element of the boundary now has the 4-velocity
\be
\label{u}
u^a= %\frac{1}{\sqrt{E}} 
\left( 1 ; \sqrt{(E-1)+\frac{2 m}{r}},0,0
 \right),
\ee
and, for an arbitrary vector in the surface $\tau=$constant, given by
$n^a = \left( 0; n^r, n^{\theta},n^{\phi} \right)$, it can be seen that
\be
u^a n_a=0.
\ee
The surfaces of constant $\tau$ are therefore orthogonal to all
 infalling boundaries that  satisfy (\ref{tl}).  We will use $\tau$ as
 our `cosmological time\footnote{We use the letter $\tau$, as this is
 proper time of observers following trajectories given by (\ref{u}).
 It is not to be confused with the `conformal time' coordinate often
 used in cosmology.}'.
%, and in everything that follows all
% tensor manipulations will be performed with respect to the geometry
% specified by $d\hat{s}^2$.

\begin{figure}[t]
\center \epsfig{file=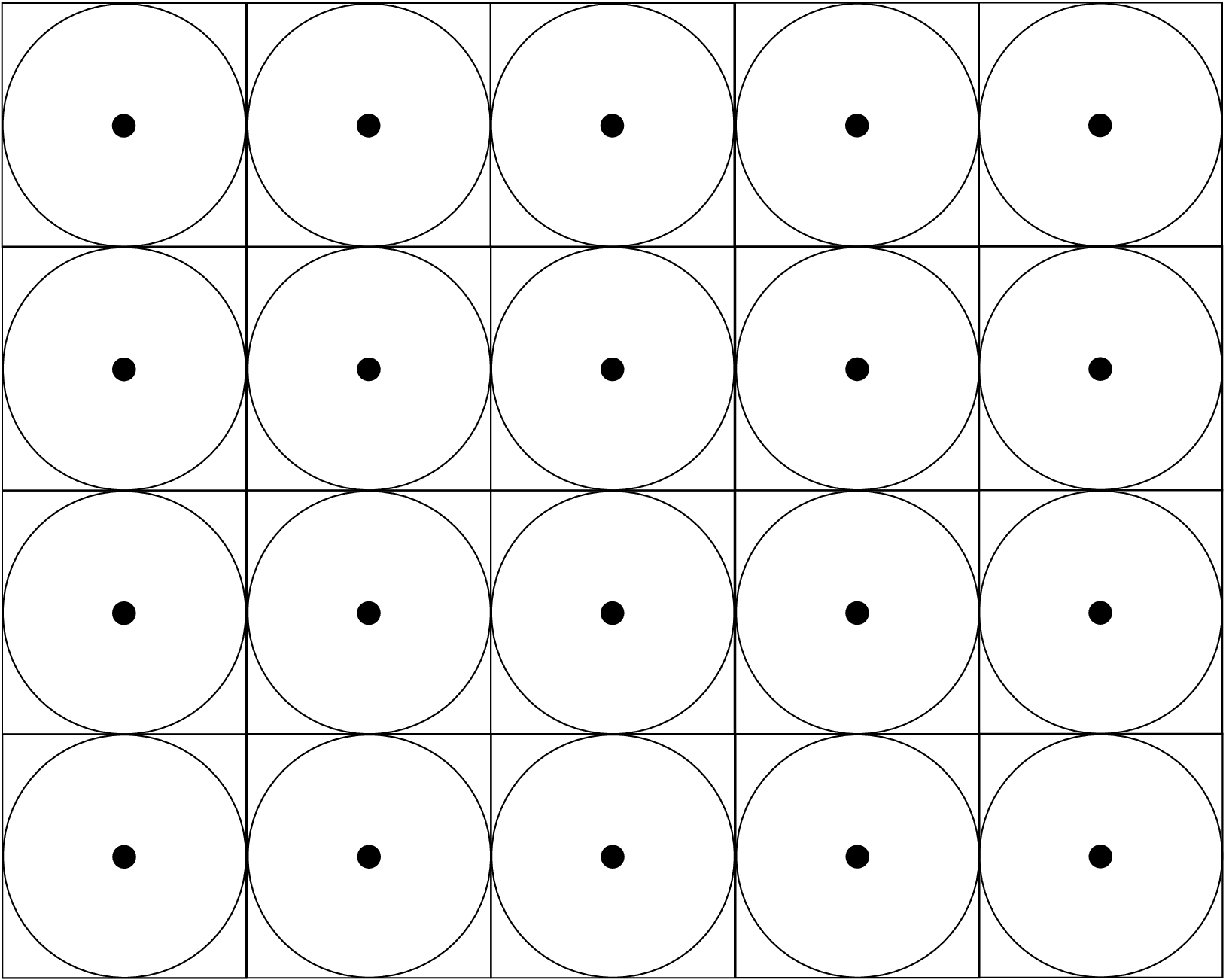,height=12cm} 
\caption{{\protect{\textit{An example of the process of replacing
	regular polytopes with $n$-spheres in a
	flat 2-space with square tiling.  The spheres touch at the
        centers of the cell faces, while at the corners of the squares
        there is a ``no man's land'' that is not covered by the spheres.
%, while at the centre of the edges of the
%squares there is an overlap between the spheres of neighbouring masses.
}}}}
\label{squares}
\end{figure}

We can now consider how to construct a lattice from individual cells
with the geometry specified by (\ref{pan}), and spherical boundaries
at $r=a(\tau)$ that satisfy (\ref{tl}).  To do this we must first
choose one of the regular lattices specified in Table \ref{table1} of
Appendix A, at
some initial time $t_i$.  A mass, $m$, must then be placed at the
centre of each cell.  This prescription leads to the highly symmetric
lattice structures that we wish to study here\footnote{One could be less
prescriptive about the symmetries involved in tessellating the 3-space,
leading to more general situations.  We will discuss this briefly
later on, but will postpone a detailed investigation for future studies.}.

We now want to make a Wigner-Seitz-like approximation, following LW,
and replace the polyhedron lattice cells with spheres.  Generalising
`Condition II' of \cite{LW}, we choose to specify these spheres as
occupying the same amount of spatial volume as the original polyhedra in the
constant curvature space in which the lattice is defined\footnote{This
is equivalent to the Condition II of \cite{LW} on lattices constructed
in positively curved 3-spaces: They will occupy $1/N$ of the
total solid angle on the hypersphere.}.  This will require the
spheres from neighbouring cells to touch at the centre of the faces
of the original lattice cells, while leaving a ``no man's land'' at
their corners (a region external to all spheres).  This type of
situation is illustrated using the square tiling of a flat 2-space in
Figure \ref{squares}.

If we now replace the geometry inside each cell by (\ref{pan}) then
the spheres will have a motion specified by (\ref{tl}), and the
orthogonality requirements we considered above will lead to tangency
of space-like surfaces of constant $\tau$ at the points where the spheres intersect.
Away from these points, in the `no man's land' region, this tangency will no longer exist.  
%It will, however, be broken in an
%opposite sense in each of the two types of regions, and so will, in
%some sense, be satisfied `on average'.

The dynamical properties of these space-times can now be determined.

The case of a space-time containing a cosmological constant is
investigated in Appendix B.

\subsection{Cosmological evolution}

From Equation (\ref{tl}) it can be seen that for $E<1$ the boundary
spheres will reach a maximum of expansion, at $a_{max}=2m/(1-E)$, and
then begin to recollapse.  For $E>1$, however, the spheres will
expand eternally, and reach infinity with a velocity $da/d\tau>0$.
The dividing case, $E=1$, corresponds to shells that just reach
infinity, and hence to the escape velocity.  Clearly this behaviour is
analogous to the behaviour of FRW cosmology, with
closed, open, and flat spatial curvatures, respectively.  This analogy
is made explicit if one re-labels variables such that 
$m\rightarrow M/2$ and $E\rightarrow 1-k$.  In this
case (\ref{tl}) becomes
\be
\frac{\dot{a}^2}{a^2} = \frac{M}{a^3}-\frac{k}{a^2},
\ee
which is obviously just the Friedmann equation, (\ref{Fried}), with
$\rho=3 M/8 \pi a^3$.  Henceforth, we will use $a$ to label the
position of the boundary sphere, and $r$ as our radial coordinate
inside each cell.  The volume of the lattice cells at future times
is then given by the corresponding volume of the bounding spheres.

We can make contact with the LW treatment by noting that in the case
of a closed space-time the LW variable $\psi$ can be
related to $E$ via $E=\cos^2 \psi$.  Clearly, $\cos^2 \psi<1$, and so
they found their space to reach a maximum of expansion, at
the value found above, before recollapsing.  The generalisation of this to
hyperbolic space is then $E=\cosh^2 \psi$, which is always $>1$, leading
to the hyperbolic expansion found above.

As in the study of LW, we find that all of the models we consider are
governed by evolution equations with identical functional form to
FRW space-times.  It was then shown in \cite{LW} that for the case of
spatially closed universes, while the evolution equation had
the same form, the scale of the problem did not: The maximum of expansion
was different in the two cases.  Here we will focus on the case of a spatially flat
universe, with $E=1$.  The expansion is then scale-invariant, and so we do not anticipate
any such discrepancies between the discrete and continuous cases to occur.

Now, while the LW construction only considered surfaces orthogonal to
the trajectory of the boundary, in
the neighbourhood of the bounding sphere,
the space-like surfaces $\tau=$constant, in the coordinate system
(\ref{pan}), are orthogonal to all shells in radial free-fall that obey
(\ref{tl}).  This allows us to extend the meshing coordinates away
from just the boundary region, and gives us a time coordinate that we
can consider to be the analogue of cosmological time in perfect fluid
cosmology.  The idea of the single bounding sphere in LW is here
replaced by a continuous family of shells\footnote{We still keep the  concept of a single
  `special' shell at $r=a$ that we will use to determine the volume
  of the cell.} following a congruence of
time-like geodesics specified by (\ref{u}), and which are identical to each other, up to a
translation in $\tau$.  Such a coordinate system will be very useful
in determining how time-like geodesics should pass between
cells.  This is particularly true in the spatially flat universe, in which
case the surfaces of constant $\tau$, in (\ref{pan}), are simply the
Euclidean 3-spaces in which the lattice cells are defined\footnote{For
the closed universe models, with $E<1$, the coordinate system given in
(\ref{pan}) has the problem that it does not cover all of the `no man's
lands' in the vicinity of the maximum of expansion.  In this case a
more complicated coordinate system can be found, as shown in
\cite{LW}.}.  We will concentrate
on this case in the analysis we perform in later sections.

\subsection{Validity of the model}

The model we are considering, as outlined above, has features that
appear at first glance to be simultaneously more realistic and less
realistic than standard FRW cosmology.  The {\it raison d'\^{e}tre} for
this model is that it does not {\it a priori} assume that discrete objects can
be simply approximated by a continuous energy density; it explicitly
maintains the discrete nature of the objects, and contains the
continuous approximation as a limit.  In this regard we consider it to
be a considerable improvement on the more usual perfect fluid description.  

Having said this, there are also obvious draw-backs.  The model itself
relies on certain approximations, such as the space-time being only an
approximate solution of Einstein's equations.  In the electro-magnetic
analogy these approximations have been shown to be
well justified \cite{WSexact1,WSexact2}.  Similar tests suggest this
is also true in the general relativistic case \cite{next}, though
further study is needed to determine this conclusively.

Beyond this, one may also question the validity of
approximating the matter content of the Universe as identical spherically
symmetric mass distributions that are equally spaced on a regular lattice.
At first glance this is clearly not true of our Universe.  Whatever
unit of structure we conceive of, it certainly is not
arranged on a regular lattice. Detailed surveys of the distribution of
galaxies have revealed
an intricate tapestry of nodes, clusters, filaments and walls that seem to have resulted 
from a stochastic process of structure formation. In fact, it has been shown that the
morphology of these structures is fractal over a wide range of scales \cite{fractal1}-\cite{fractal4}. 
This suggests that a less symmetric structure than a regular
lattice would be more realistic.  However, we wish to retain as much as possible of the cosmological
principle\footnote{A statistical cosmological principle (in which
one assumes statistical homogeneity and isotropy) would be more desirable, but is currently
impossible to implement. The next best thing appears to be a regular
lattice.} in our study, and it makes sense
to consider the simplest, most symmetric case first.  

Now let us return to the question of the masses.  Again, most objects
in the Universe do not appear to be identical and perfectly
spherical.  Never the less, it is certainly not unusual to approximate
the gravitational fields of non-spherical objects (such as disc galaxies)
as being spherical, and taking equal mass objects seems like a
justifiable approximation in order to make progress in
understanding the problem at hand.  Taking into
account the detailed shape and structure of every object in the
Universe would obviously be prohibitively difficult.  One would hope
that future studies would make progress from the simplest case studied
here, to more realistic situations.

%And what are our fundamental units of mass? For a start it is straightforward to work out
%the relationship between the various length scales. The background cosmology obeys
%$a(\tau)=(\tau/\tau_0)^{2/3}$ and let us parametrise $L_0$ such that $H_0L_0=\alpha_0$. We
%then have
%\begin{eqnarray}
%\alpha_0=\left(\frac{mH_0}{4}\right)^{\frac{1}{3}}=\left(\frac{M}{M_{MW}}h\right)^{1/3}\times 10^{-4}
%\end{eqnarray}
%where $M_{MW}$ is the mass of the Milky Way. An obvious choice is to consider Milky Way
%mass  galaxies laid regularly out on a lattice and we find the lattice spacing is a few Mega Parsecs.
%Clusters of galaxies will make lattice spacing fractionally larger but not by much. A radical alternative
%is to consider dark matter particles with masses of a few 100 GeV- the lattice spacing reduces to
%a few metres on the side.

An obvious choice for the central masses in this model is to consider
them to be galaxies with the same mass as the Milky Way.  In this case
we find the spacing between masses should be of the order of a few
mega-parsecs.  Clusters of galaxies would be another sensible choice,
and would correspond to a fractionally larger spacing of
masses. Alternatively, one might also consider dark matter particles
(with masses of a few 100 GeV).  Such a choice may be more suitable
for applying the model considered here to the early Universe.  The
lattice spacing would then reduce to a few metres.

Throughout this paper we will use Milky Way type masses, although
we have also considered other objects and found our results to be
largely insensitive to this choice (unless we consider very large
masses).

\section{Photon Trajectories in a Lattice}
\label{3}

We will now consider the trajectories of null particles in this
space-time.  These will be of basic importance for understanding the
optical properties of the lattice universe. All analytic results
presented here will be backed up by numerical analysis.

\subsection{Null geodesic equations}

The Euler-Lagrange equations derived from the line-element (\ref{pan}) are
\bea
\frac{d}{d\lambda} \left( \left(1-\frac{2 m}{r} \right) \dot{\tau} +
\sqrt{E-\left( 1-\frac{2 m}{r} \right)} \dot{r}
\right) &=& 0\\
\frac{d \dot{r}}{d\lambda}
- \sqrt{E-\left( 1-\frac{2
   m}{r}\right)} \frac{d\dot{\tau}}{d\lambda} &=&
-\frac{m}{r^2} \dot{\tau}^2
+E r \dot{\theta}^2+E r \sin^2 \theta \dot{\phi}^2 \\
\frac{d}{d\lambda} \left( r^2 \dot{\theta} \right) &=& r^2 \sin \theta \cos
\theta \dot{\phi}^2 \\
\frac{d}{d\lambda} \left( r^2 \sin^2 \theta \dot{\phi} \right) &=& 0,
\eea
together with the null constraint
\be
\label{constraint}
-\left(1-\frac{2 m}{r} \right) \dot{\tau}^2+ \dot{r}^2 +E r^2
\dot{\theta}^2+E r^2 \sin^2 \theta \dot{\phi}^2-2 \sqrt{E-\left(
 1-\frac{2 m}{r}\right)} \dot{r} \dot{\tau} =0,
\ee
where dots indicate derivatives with respect to $\lambda$, an affine
parameter along the geodesic.  Integrating these equations gives

\bea
\left(1-\frac{2 m}{r} \right) \dot{\tau} +
\sqrt{E-\left( 1-\frac{2 m}{r} \right)} \dot{r} &=& B\\
\dot{\phi} &=& \frac{J_{\phi}}{r^2 \sin^2 \theta}\\
\dot{\theta}^2 &=& \frac{J^2}{r^4}- \frac{J_{\phi}^2}{r^4 \sin^2\theta}\\
\dot{r}^2 &=& \frac{B^2}{E}-\frac{J^2 }{r^2} \left( 1-\frac{2 m}{r} \right),
\eea
where $B$, $J$ and $J_{\phi}$ are constants.
Clearly, one could rotate coordinates so that $\theta = \pi/2$ in each
cell, although this will not be as useful here as it usually is, as we
will want to match different coordinate systems between cells.
Rotating them differently in each cell would confuse things.

The geodesic equations can also be found in Cartesian coordinates (when
$E=1$), which are useful for numerical implementation, and in the presence of a cosmological constant.  The
equations for these cases are given in Appendices C and D, respectively.

\subsection{Matching trajectories at boundaries}

Now consider the following situation:  A photon is emitted in one lattice
cell and passes through a cell boundary (or, more likely, several cell boundaries),
before it is observed.  In order to make predictions about
observations of events that occur in cells that are at some distance
from the cell inhabited by the observer, we must be able to propagate
photon trajectories between cells.

One may initially suspect that at the boundary between cells it may be
suitable to simply perform a transformation of spatial coordinates
from one cell to another, via a suitable translation of the origin of
the coordinate system.  However, it is soon seen that this is {\it
  not} a viable way of propagating null trajectories between cells.
To see why consider the null constraint equation, (\ref{constraint}),
and a photon that is following a radial trajectory in both the first
and second cells.  If we denote the spatial coordinates of the first
cell by un-hatted coordinates, then (\ref{constraint}) gives
\be
\nonumber
-\left(1-\frac{2 m}{{r}} \right) \dot{\tau}^2+ \dot{{r}}^2 
-2 \sqrt{E-\left(1-\frac{2 m}{{r}}\right)} \dot{{r}} \dot{\tau} =0,
\ee
where $\dot{{r}}>0$, as the photon is leaving the cell.  If one
were to perform the simple translation discussed above, keeping
$\dot{\tau}$ the same, then the
equation above would transform to
\be
\nonumber
-\left(1-\frac{2 m}{\hat{r}} \right) \dot{\tau}^2+ \dot{\hat{r}}^2 
+2 \sqrt{E-\left(1-\frac{2 m}{\hat{r}}\right)} \dot{\hat{r}} \dot{\tau} =0,
\ee
as $\dot{{r}}=-\dot{\hat{r}}$ and ${r}=\hat{r}$ at the boundary.  Hats
denote the spatial coordinates of the second cell.  It can immediately be seen that
this new equation is not compatible with the null constraint,
(\ref{constraint}), in the second cell, as it has the wrong sign
before the third term on the left-hand side.  We must therefore be more
careful.

Our basic criterion for matching trajectories across boundaries is
that observable quantities, such as photon frequencies and directions,
should be independent of which coordinate system an observer moving with the
boundary chooses to use.

Now, at the boundary, it will only be a set of measure zero
trajectories that actually pass through the single shell that is described by
our `special' bounding sphere, which prescribes the volume of the
cell.  All other trajectories will pass through the ``no man's land'',
that is outside of all bounding spheres.
%All other trajectories will either pass though in the `overlap' region, which is
%the intersection of the interiors of the two bounding spheres, or in
%the `no man's land', that is outside of all bounding spheres.  
This means that observers comoving with the family of shells that obey
(\ref{u}) will, in general, be in relative motion with respect to the
cell boundary when the photon in question passes them.  Such motion
should be expected to result in a redshift between comoving observers
from neighbouring cells who are both at the same point on the cell
boundary at the same time.  We will now consider two different methods of
accounting for this effect.

In Method I we will appeal to the approximate tangency between the space-like volumes
of constant $\tau$.  A simple, approximate matching criterion is then
given by the condition that $\dot{\tau}$ is the same on leaving one
cell, as it is on entering the next.  This will be shown in subsequent
section to correspond to the condition that, at the boundary, the
frequency of a photon measured by a comoving observer from the
first cell should be identified with the frequency measured by comoving
observers from the second.  Clearly, this criterion will
not be exactly satisfied by all trajectories\footnote{In fact, any that do not
pass through the `special' bounding sphere at the boundary.}, but it
may be satisfied in an approximate way over many trajectories.
%Those that pass through the boundary in the `overlap' region will pick up a
%change in $\dot{\tau}$ with the opposite sign to those that pass
%through in the `no man's land'.

In Method II we will attempt to account for the relative motion between
comoving observers from different cells in a more detailed way.
Instead of assuming that the effects described above cancel
approximately, we will work out the redshift between a comoving observer who
is momentarily at the cell boundary, and an observer who is moving
non-geodesically with the boundary.  Rather
than identifying the frequency measured by comoving observers from
each cell, we will then identify the redshift
measured by the observers who are moving with the
boundary.  This method will be more complicated, but we expect it to
account for the imperfect tangency in a more complete way.

The results of using Method II will turn out to be very similar to the results of
using Method I.  This provides us with motivation for considering the concept of
approximate tangency between the space-like surfaces of neighbouring cells
to be a valid one.

\subsubsection{Method I: A simple, approximate matching}

Using the matching condition that $\dot{\tau}$ on leaving the first
cell is the same as $\dot{\tau}$ on entering the second cell, we are
left with the task of finding $\dot{r}$, $\dot{\theta}$ and
$\dot{\phi}$ in the new cell.  To do this consider the following
decomposition of the 4-vector tangent to the null geodesic,
\be
k^a=\frac{dx^a}{d\lambda} = \left( \dot{\tau};
\dot{r},\dot{\theta},\dot{\phi} \right) = (-u^b k_b) (u^a+n^a),
\ee
where $u^a$ is the trajectory of an observer on one of the free
falling shells specified by (\ref{u}).  The 4-vector $n^a$ is the same
unit space-like vector that was considered below Equation (\ref{u}).

Now consider describing the null geodesic in a neighbourhood of the
observer using the coordinate system of the first cell.  This results
in
\be
k^a = \dot{\tau} \left( 1; \sqrt{(E-1)+\frac{2 m}{r}}+n^r,
n^{\theta}, n^{\phi} \right),
\ee
where we have made use of the fact that $u^ak_a=-\dot{\tau}$, and
($n^r$,$n^{\theta}$,$n^{\phi}$) obey the normalisation condition
\be
\frac{(n^r)^2}{E}+r^2 (n^{\theta})^2+r^2 \sin^2 \theta (n^{\phi})^2=1.
\ee
Now consider the same geodesic, in the same neighbourhood of the same
observer, but this time using the spatial coordinate system of the second
cell.  In this case the same reasoning gives the same tangent vector as
\be
\label{kn}
k^{\hat{a}} = \dot{\hat{\tau}} \left( 1; \sqrt{(E-1)+\frac{2 m}{\hat{r}}}+n^{\hat{r}},
n^{\hat{\theta}}, n^{\hat{\phi}} \right),
\ee
where hats denote indices in the second coordinate system (that of the
new cell into which the photon is propagating).  As discussed above, the cell boundary is
equidistant to the central mass of each cell, so that $r=\hat{r}$, and the vector
$n^{\hat{a}}$ obeys a similar normalisation condition to that of
$n^a$. 

Method I now tells us that $\dot{\tau}$ in the neighbourhood of the observer
with 4-velocity (\ref{u}) at the boundary is, approximately, common to both coordinate systems, so that 
\be
\dot{\tau} \vert_{\text{out}} \simeq \dot{\hat{\tau}}\vert_{\text{in}}.
\ee
We can then relate $\left(\dot{r},\dot{\theta},\dot{\phi} \right)$ to
$\left( \dot{\hat{r}},\dot{\hat{\theta}},\dot{\hat{\phi}} \right)$ using the
condition that the projection of the tangent vector of the null
geodesic into the rest space of an observer at the boundary should be
independent of which spatial coordinate system is being used.

To see how this works consider first a radial geodesic.  The
condition $\dot{\theta}=\dot{\phi}=0$ then gives
$n^{\theta}=n^{\phi}=0$, and similarly in the hatted coordinates.  We
also have that $n^r=1$ in the coordinates of the first cell (this says
that the geodesic is outgoing).  In the coordinates of the second
cell, however, we have $n^{\hat{r}}=-1$, as the geodesic is ingoing in the new
coordinate system (i.e. moving to smaller $r$).  We can then write the
relation between $\dot{r}\vert_{\text{in}}$ and $\dot{r}
\vert_{\text{out}}$ as
\be
\dot{r} \vert_{\text{out}} = -\dot{r}\vert_{\text{in}} + 2 \dot{\tau}
\sqrt{(E-1)+\frac{2 m}{r}}.
\ee
This satisfies the constraint equation exactly, and we avoid the
problem noted above.

Now consider a general (not necessarily radial) geodesic.  In this
case we have that the expression for $k^a$ above gives, for $E=1$,
\bea
\label{b1}
\dot{\theta} &=& \dot{\tau} n^{\theta}\\
\label{b2}
\dot{\phi} &=& \dot{\tau} n^{\phi}\\
\label{b3}
\dot{r} &=& \dot{\tau} n^r + \dot{\tau} \sqrt{\frac{2 m}{r}}.
\eea
Similar relations are obeyed by the hatted coordinates.  The
expression $k^a k_a=0$ now gives
\be
k^a k_a = k^a k^b g_{ab} = (n^a n^b g_{ab}-1) \dot{\tau}^2=(n^{\alpha}
n^{\beta} \delta_{\alpha \beta}-1) \dot{\tau}^2,
\ee
where Greek indices run over spatial coordinates, and $\delta_{\alpha
  \beta}$ represents the spatial metric of the Euclidean 3-space. This clearly
satisfies the null condition, as $n^an_a=1$.  The constraint equation
can then be seen to be valid in any two coordinates systems related by
a transformation of spatial coordinates of the form  $x^{\hat{\alpha}}
= {\Lambda^{\hat{\alpha}}}_{\alpha} x^{\alpha}$, where ${\Lambda^{\gamma}}_{\hat{\alpha}}
{\Lambda^{\hat{\alpha}}}_{\alpha}= {\delta^{\gamma}}_{\alpha}$.

The procedure for propagating a general geodesic through the boundary
using Method I is then the following:  

\vspace{10pt}

(1) Take $\dot{\tau}, \dot{r},\dot{\theta}$
and $\dot{\phi}$ from the first cell and use them to find $n^r,
n^{\theta}$ and $n^{\phi}$,
using the relations (\ref{b1})-(\ref{b3}).  

\vspace{10pt}

(2) Transform the ($r$, $\theta$, $\phi$) coordinate system into a system with the same symmetries as the cell
(this is Cartesian coordinates for $E=1$).  Translate the origin of
the coordinates from the central mass of one cell, to the central mass
of another.  Transform to a new set of spherical coordinates, $\left(
\hat{r}, \hat{\theta}, \hat{\phi} \right)$.  Calculate $n^{\hat{r}},
n^{\hat{\theta}}$ and $n^{\hat{\phi}}$ in these new coordinates.  The
vector $n^{\hat{a}}$ is then given in terms of $n^a$ by the transformation
\be
n^{\hat{a}} = \frac{\partial x^{\hat{a}}}{\partial x^a} n^a,
\ee
where the two coordinate systems are related, for $E=1$, by
\bea
\hat{r}^2 &=& r^2 +x_0^2 -2 r x_0 \cos \phi \sin \theta\\
\cos^2 \hat{\phi} &=& \frac{(r \cos \phi \sin \theta
 -x_0)^2}{(x_0^2+r^2 \sin^2 \theta -2 r x_0 \sin \theta \cos \phi)}\\
\cos^2 \hat{\theta} &=& \frac{r^2 \cos^2 \theta}{(r^2 +x_0^2 -2 r x_0
 \sin \theta \cos \phi)},
\eea
and $x_0$ is given by the translation between Cartesian coordinate
systems, $\hat{x} = x-x_0$.  This will be equal to the width of a
cell, at the time the photon hits the boundary.

\vspace{10pt}

(3) Use these values to calculate $\dot{\hat{r}},\dot{\hat{\theta}}$ and
$\dot{\hat{\phi}}$ using (\ref{b1})-(\ref{b3}), given that
$\dot{\hat{\tau}}= \dot{\tau}$.  These are then the initial
conditions for propagating the null geodesic through the new cell.

\vspace{10pt}

(4) Repeat at the next cell boundary.

\vspace{10pt}

\subsubsection{Method II: A more elaborate, and accurate matching}

The above method did not take into account the fact that along individual
trajectories the observers with 4-velocity $u^a$ on either side of a
cell boundary can have a relative
velocity between them.  This is due to the imperfect tangency between
surfaces of constant $\tau$ on either side of a cell boundary.  
To calculate the effect of this we will
consider an observer moving (non-geodesically) with the boundary.  The
4-velocity of such an observer, $w^a$, can be given by
\be
w^a = (w^{\tau} ; w^r ,0 ,0),
\ee
and satisfies $w^a w_a = -1$, if the observer is time-like, and
radially moving away from the mass at the centre of each cell.  The normalisation
condition gives us that
\be
w^{\tau} = \frac{\sqrt{1-\frac{2 m}{r}+(w^r)^2}-\sqrt{\frac{2 m}{r}}
 w^r}{(1-\frac{2 m}{r})}.
\ee
The 3-velocity, $w^r$,  can then be worked out straight-forwardly. For
$E=1$ we can use Euclidean geometry to give $w^r$ in terms of the expansion of the shell that
defines the volume of the lattice cell, $\dot{a}$, as
\be
\label{wr}
w^r = 
%\left( \frac{\pi}{6} \right)^{1/3} 
\sqrt{1+\left(  \delta
 x\right)^2}\dot{a},
\ee
where $\delta x$ is the distance from the centre of the face of the
particular lattice cell that is being crossed, as a fraction of the
distance from the cell centre to the centre of the cell face.

The frequency of a photon with tangent 4-velocity $k^a$ measured by this observer is then given by
\be
\label{z1b}
-w^a k_a = \dot{\tau}\sqrt{1- \frac{2 m}{r} + (w^r)^2} -
\frac{(w^r-\sqrt{\frac{2 m}{r}} \sqrt{1-\frac{2 m}{r} +(w^r)^2})}{(1-
 \frac{2 m}{r})} \dot{r}.
\ee
The values of $\dot{\tau}$ and $\dot{r}$ are given by the
solutions to the null geodesic equations stated above.
Similarly, in the second cell we have
\be
\label{z3b}
-w^{\hat{a}} k_{\hat{a}} = \dot{\hat{\tau}}\sqrt{1- \frac{2 m}{r} + (w^r)^2} -
\frac{(w^r-\sqrt{\frac{2 m}{r}} \sqrt{1-\frac{2 m}{r} +(w^r)^2})}{(1-
 \frac{2 m}{r})} \dot{\hat{r}},
\ee
where, by symmetry, the 4-velocity of the observer in the cell
boundary is the same for both cells.  We now wish to identify the
frequency of the photons measured by the observers moving with
the boundary, giving the condition $-w^a k_a=-w^{\hat{a}} k_{\hat{a}}$.

If, as before, we now use the projection of the 4-vector $k^a$ into the
Euclidean 3-space, then we have that $\dot{\theta}$, $\dot{\phi}$ and
$\dot{r}$ in the first cell are again given by Equations
(\ref{b1})-(\ref{b3}).  Similar expressions are satisfied by $\dot{\hat{\theta}}$,
$\dot{\hat{\phi}}$ and $\dot{\hat{r}}$, but with hatted coordinates.

We can now use (\ref{z1b}) and (\ref{z3b}) to calculate $\dot{\hat{\tau}}$ in terms of known
quantities, as
\be
\label{tauhat}
\dot{\hat{\tau}} = \frac{\dot{\tau} \sqrt{1- \frac{2 m}{r} + (w^r)^2} -
\frac{(w^r-\sqrt{\frac{2 m}{r}} \sqrt{1-\frac{2 m}{r} +(w^r)^2})}{(1-
 \frac{2 m}{r})} \dot{r}}{\sqrt{1- \frac{2 m}{r} + (w^r)^2}-
 \frac{(\hat{n}^r +\sqrt{\frac{2 m}{r}})}{(1-\frac{2 m}{r})}  \left(
 w^r-\sqrt{\frac{2 m}{r}} \sqrt{1-\frac{2 m}{r} +(w^r)^2}\right)}.
\ee
For $w^r = \sqrt{2m/r}$ we recover $\dot{\hat{\tau}}=\dot{\tau}$, as
used in Method I.  More generally, the value of $\dot{\hat{\tau}}$
given by (\ref{tauhat}) can be used, together with
$n^{\hat{\theta}}$, $n^{\hat{\phi}}$ and $n^{\hat{r}}$, to find
$\dot{\hat{\theta}}$, $\dot{\hat{\phi}}$ and $\dot{\hat{r}}$.
The steps to be followed at the boundary using Method II are therefore:

\vspace{10pt}

(1) Find $w^r$ at the boundary, using (\ref{wr}).

\vspace{10pt}

(2) Take $\dot{\tau}, \dot{r},\dot{\theta}$
and $\dot{\phi}$ from the first cell and use them to find $n^r,
n^{\theta}$ and $n^{\phi}$,
using the relations (\ref{b1})-(\ref{b3}).  

\vspace{10pt}

(3) Find $n^{\hat{a}}$ in terms of $n^a$ via the transformation
\be
n^{\hat{a}} = \frac{\partial x^{\hat{a}}}{\partial x^a} n^a,
\ee
where the two coordinate systems are related by the same coordinate
transformation as in Method I.

\vspace{10pt}

(4) Find $\dot{\hat{\tau}}$ using (\ref{tauhat}).

\vspace{10pt}

(5) Use the values found in (3) and (4) to calculate $\dot{\hat{r}},\dot{\hat{\theta}}$ and
$\dot{\hat{\phi}}$ using (\ref{b1})-(\ref{b3}) (with hats added).  These are then the initial
conditions for propagating the null geodesic through the new cell.

\vspace{10pt}

(6) Repeat at the next cell boundary.

\vspace{10pt}

%We will find in later sections that the difference between using
%Method I and Method II is actually very small, lending credence to the
%idea of an `average' tangency between the space-like surfaces of
%constant $\tau$ at the boundaries between cells, as discussed above.

\subsection{Numerical Implementation}

We have implemented the methods described above numerically, so that we can propagate
geodesics with any initial positions and directions out to arbitrarily large
distances. This is effectively the equivalent of the ray tracing methods applied in
simulations of lensing through cosmological configurations. 
Note that in all plots, we integrate back in time from the observer's
current position, and fix the affine
parameter, $\lambda$, to be initially zero. 

A few brief comments on this seem in order at this point.  
We find that while integrating the geodesic equations through a Schwarzschild cell
it is preferable to use Cartesian coordinates (as given in Appendix C),
rather than the natural, spherical coordinates of the Schwarzschild
geometry. In this way we can avoid coordinate
singularities (where, for example, $\sin\theta=0$).  Although these singularities are rare for
any single geodesic, they can still occur occasionally if one
integrates over many different geodesics in order to obtain statistically significant results
from a Monte Carlo simulation.

A further point to comment on is that with our choice of Milky
Way sized masses, the ratio between the Schwarzschild radius and the cell
size is minute (of order $10^{-8}$). This means that large deflection
events (as trajectories pass nearby the central mass) are almost
non-existent, and that all but a few geodesics suffer only negligible
deflections as they traverse a single Schwarzschild
cell. With this fact in hand, we find it sufficient to consider
rectilinear trajectories (akin to the Born approximation), and to
calculate our optical quantities along them.

\section{Cosmological Redshift From Schwarzschild Patches}
\label{4}

Now that we have the equations for null geodesics, we can calculate
the redshifts between source and observer that are so important in cosmology.
The redshift will, of course, depend on the motion of the
source and observer.  These can, in principle, be completely
arbitrary.  However, the closest analogy to a comoving source and
observer in FRW cosmology will be a source and observer that are
comoving with one of the family of free-falling shells with 4-velocity
specified by (\ref{u}).  As always, the redshift,
$1+z$, is given as the ratio of the frequency, $-u^ak_a$, measured at the source and
observer\footnote{This is the case for both Methods I and II.  The
  4-velocity of the non-geodesic observers in Method II, $w^a$, is
  only used for propagating photons across boundaries.}.  Here this is
\be
- u^a k_a =  \dot{\tau}
\ee
so the redshift is
\be
1+z = \frac{\dot{\tau} \vert_e}{ \dot{\tau} \vert_o},
\ee
where subscript $e$ and $o$ denote when the photon was emitted and
observed, respectively.  In general this quantity will need to be
calculated numerically.  We find, however, that we are able to deduce
reasonably good analytic approximations to the numerical results.  We
will present our analytic approximations first, and then proceed to
compare them to numerical solutions.

\subsection{Analytic approximation}

To find analytic approximations for the redshift consider first a single null trajectory.  For an expanding
space-time we then have from the geodesic equations that
\be
\label{t1}
\dot{\tau} = B \frac{\left(1- \alpha \sqrt{\frac{2 m}{r}} \right) }{(1 - \frac{2 m}{r})}.
\ee
The factor of $\alpha$ is included to account for the magnitude of $\dot{r}$ as a
fraction of the total `velocity',
$\sqrt{\dot{r}^2+r^2 \dot{\theta}^2+r^2 \sin^2 \theta \dot{\phi}^2} \simeq B$, and is given by
$\alpha\equiv \dot{r}/B$ for any particular cell\footnote{For a geodesic with
$J=J_{\phi}=0$ we have $\alpha=1$, as the photon is moving entirely in
the radial direction.  More generally $\alpha \in [0,1]$.}.

\begin{figure}[t]
\center \epsfig{file=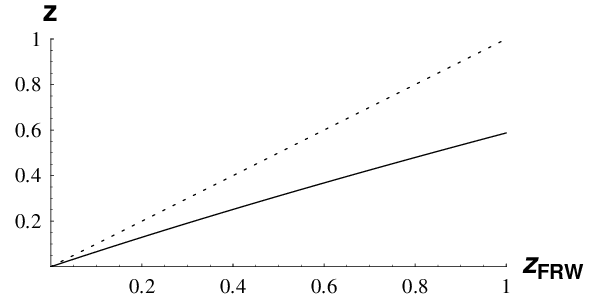,height=8cm} 
\caption{{\protect{\textit{The approximate relation expected between redshift in a
	lattice universe, $z$, and the corresponding redshift
	in an FRW universe, $z_{FRW}$.  The solid line is for null trajectories not lying on a
	principle axis, and the dotted line is for near radial geodesics.}}}}
\label{zfig}
\end{figure}

\begin{figure}[t]
\center \epsfig{file=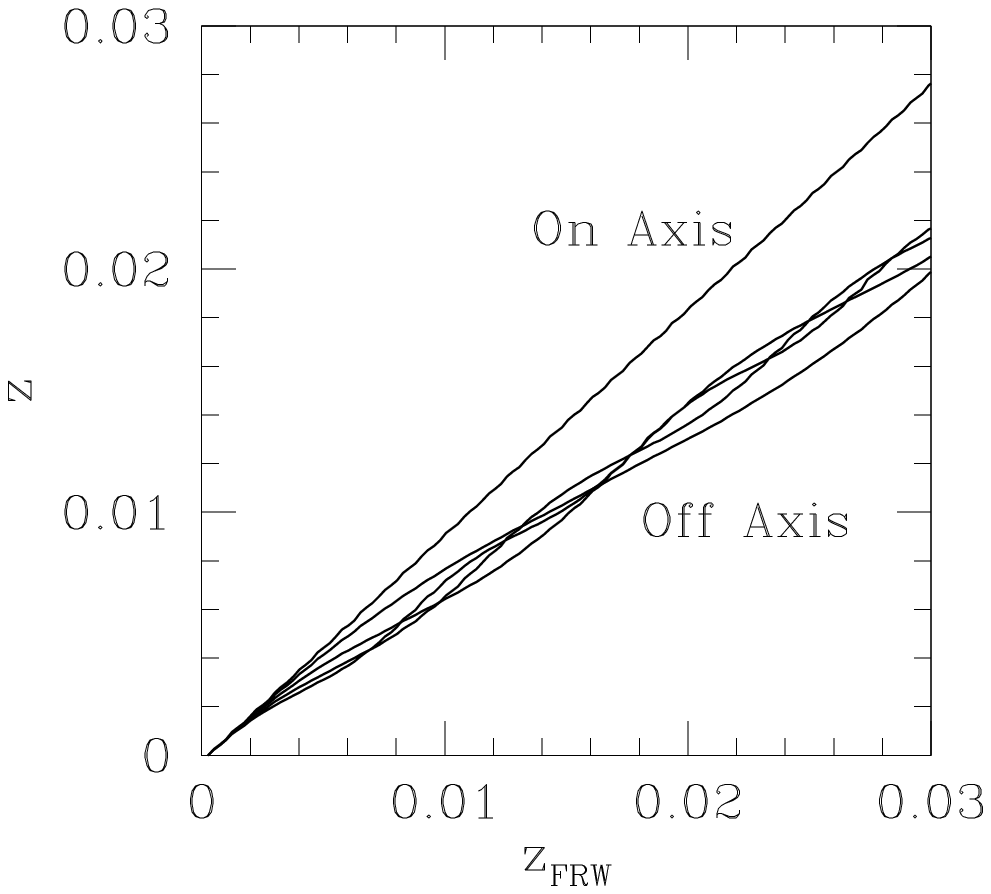,height=7.5cm} 
 \epsfig{file=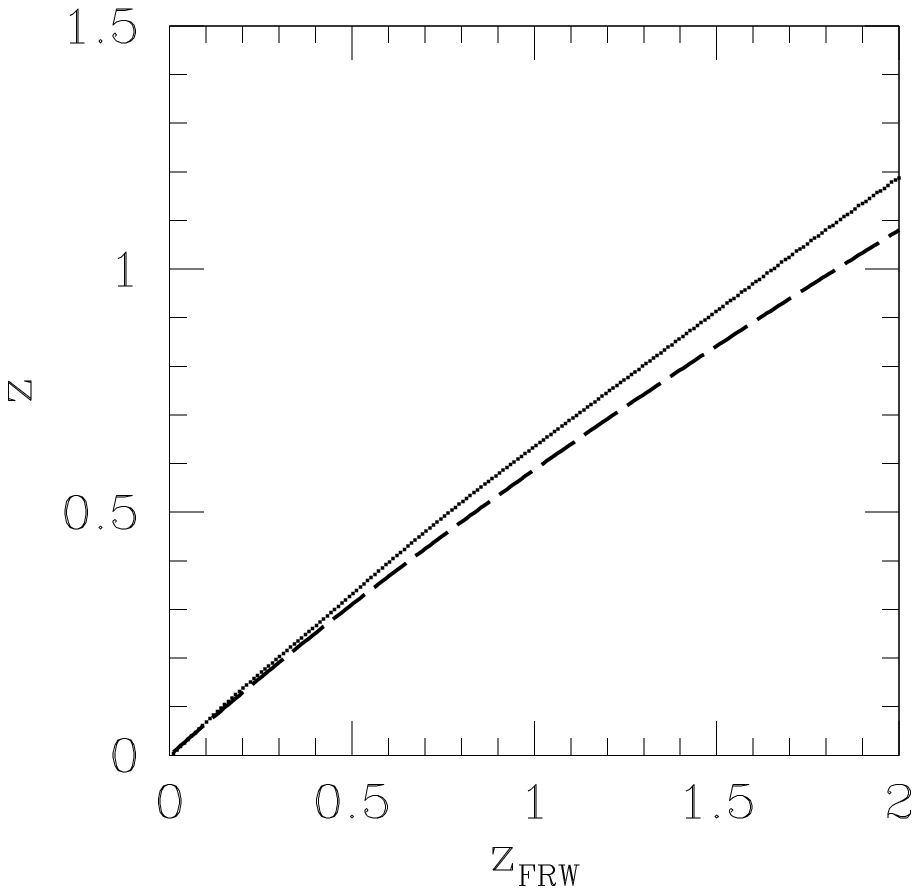,height=7.5cm} 
% \vskip -0.6in
\caption{{\protect{\textit{The relation between redshift in a
	lattice universe, $z$, and the corresponding redshift in an
	FRW universe, $z_{FRW}$, for 5 different trajectories starting
	off with different orientations. Close to the observer
	$($left panel$)$ there is still a fair amount of
	scatter, but after traversing a few Schwarzschild domains,
	this scatter becomes negligible $($right
	panel$)$. We also plot the analytic fit with
	$\langle \gamma \rangle =2/3$ from expression $(\ref{ZZFRW})$ $($dashed
	line$)$. Equation $(\ref{ZZFRW})$ with $\langle \gamma \rangle
	=7/10$ $($solid line$)$ lies directly on top of the numerical results $($dotted line$)$.}}}}
\label{Znum1}
\end{figure}

\begin{figure}[t]
\center 
 \epsfig{file=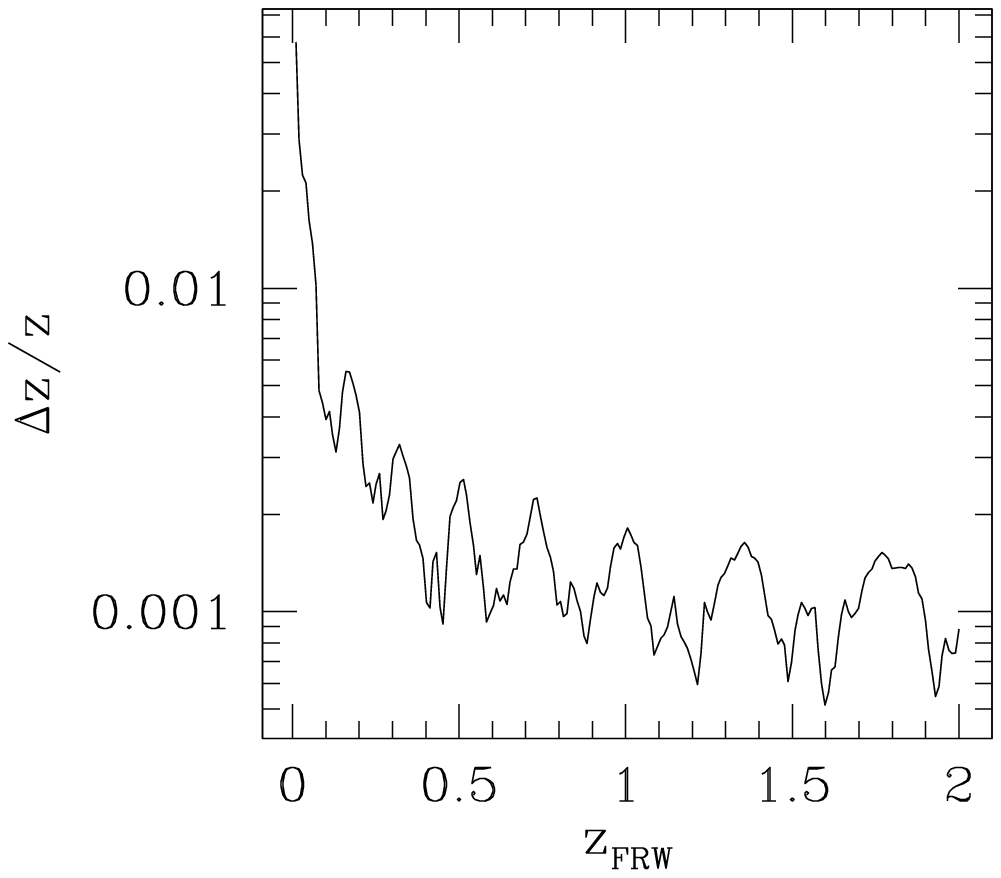,height=8cm} 
%  \vskip -0.6in
\caption{{\protect{\textit{The error in the $z-z_{FRW}$ relation for 40 random trajectories.}}}}
\label{ZStat}
\end{figure}

For $2m \ll r$ we then have
\be
\dot{\tau} \simeq B (1- v (\alpha -v)),
\ee
where $v \equiv \sqrt{2m /r}$ is the velocity of the observer.
The redshift induced by a photon travelling through a single
cell, $(1+\delta z_i)$, is then
\be
(1+\delta z_i)=\frac{\dot{\tau}
  \vert_{\text{in}}}{\dot{\tau} \vert_{\text{out}}} 
\simeq \frac{(1-v_{in} (\alpha_{in}-v_{in}))}{(1-v_{out}
  (\alpha_{out}-v_{out})} \simeq 1+ v (\alpha_{out}-\alpha_{in}),
\ee
as $B$ is constant inside each cell.  In the last equality we have
used $v_{in} \simeq v_{out}$ for a photon passing through a single
cell. For small $z$ the total redshift induced by travelling through
$n$ cells is now
\be
\label{z2}
1+z = \prod_{i=1}^{n} (1+\delta z_i) \simeq 1+2 \sum_i \alpha_i
\sqrt{\frac{2m}{a_i}}\simeq 1+2 \sqrt{\frac{2m}{a_0}} \tau_0^{1/3}
\int \frac{di}{\tau^{1/3}} \alpha_i
\ee
where $a_i$ denotes $a$ at the moment the photon enters the $i$th
cell.  

We now need $i=i(\tau)$ in order to proceed (this is the cell number that
the photon is in, as a function of $\tau$).  
%Along a radial null geodesic we
%can integrate (\ref{t1}) to get
%\be
%\tau-\tau_1 = \sqrt{r} (\pm \sqrt{r}-\sqrt{8 m}) \pm 4 m \ln
%\vert \sqrt{2 m} \pm \sqrt{r} \vert ,
%\ee
%where $\tau_1$ is a constant of integration.  We then have the time to
%cross a single cell radially is
%\be
%\delta \tau_i \simeq 2 r_i +4 m \ln \left( \frac{1+\sqrt{\frac{2
%     m}{r_i}}}{1-\sqrt{\frac{2 m}{r_i}}} \right),
%\ee
%where we have 
Assumed the expansion of the cell is slow compared to the
scale of the photon crossing time\footnote{This should be a good
approximation as long as the cells are small.}, to lowest order in $m/r$ we then have that the
crossing time of a region, $\Delta i$, is given by
\be
\frac{\Delta \tau_i}{\Delta i} \simeq - 2 a_i,
\ee
with the continuum limit
\be
\frac{d\tau}{di} \simeq -2 a_0 \left( \frac{\tau}{\tau_0} \right)^{2/3}.
\ee
More generally, for non-radial geodesics, we can write
\be
\frac{d\tau}{di} \simeq -2 \beta_i a_0 \left( \frac{\tau}{\tau_0} \right)^{2/3},
\ee
where $\beta_i$ is the distance across the cell as a fraction of the
length of the radial trajectory. Substituting this into (\ref{z2}) and integrating gives the redshift as

\bea
\label{z3}
1+z &\simeq& 1- \frac{\sqrt{2m}}{a_0^{3/2}} \tau_0 \left< \frac{\alpha}{\beta} \right> \ln \frac{\tau_e}{\tau_0}\\
&\simeq& 1- \langle \gamma \rangle \ln
  \left( \frac{\tau_e}{\tau_0} \right)^{2/3}\\
&\simeq& 1+ \langle \gamma \rangle \ln
  \left( \frac{a_0}{a_e} \right)\\
&=& 1+ \langle \gamma \rangle \ln
  \left( 1+z_{FRW} \right) \label{ZZFRWE}\\
&\simeq& (1+z_{FRW})^{\langle \gamma \rangle}, \label{ZZFRW}
\eea
where $\langle \gamma \rangle \equiv \langle \alpha /
  \beta \rangle$.  Strictly speaking, the last step here is only valid
  up to linear order in $\ln \left( 1+z_{FRW} \right)$.  This is
  perfectly acceptable, however, as the calculation before hand has only been
  performed up to this order of accuracy.  Now, if we use the
  fact that redshifts should combine as factors\footnote{Such that
  $(1+z_{1\rightarrow 3})=(1+z_{1\rightarrow 2}) (1+z_{2\rightarrow
  3})$.}, then we have good reason to suspect that (\ref{ZZFRW}) will
  be a better approximation to $1+z$ than (\ref{ZZFRWE}).  The reason
  for this is that Equation (\ref{ZZFRW}) can be seen to combine
  factors of redshift in exactly the way required, and to have the correct linear term in an expansion of $\ln
  \left( 1+z_{FRW} \right)$ around $0$.  For these reasons we
  expect it to be a reasonable non-linear extension of the relation between $z$
  and $z_{FRW}$.  This is confirmed numerically.

The angle brackets here refer to an ensemble average, over all the cells
along a trajectory.  For radial geodesics $\langle \gamma
\rangle \rightarrow 1$, and so $z\rightarrow z_{FRW}$.  For general
geodesics it is shown in Appendix E that $\langle \gamma
\rangle \simeq 2/3$.  The redshift in a lattice universe is then found
to go like
\be
\label{z1}
1+z\simeq (1+z_{FRW})^{2/3},
\ee
for a trajectory that is not aligned with any principle axis. This
result marks a significant deviation from the corresponding
observable in FRW. It is
shown graphically in Figure \ref{zfig}.

\subsection{Numerical Results}

If, instead of making approximations, we integrate the geodesic
equations numerically, then we can get more accurate results, and
determine the validity of the simple expression (\ref{z1}).
Doing this, we find that the analytical results are good (though not
perfect) approximations. As one would expect, we do find an initial scatter in the $z-z_{FRW}$
relation as the geodesics traverse the first few Schwarzschild cells.

In Figure \ref{Znum1} we plot the relationship between $z$ and $z_{FRW}$ for five trajectories with different
initial directions. The steepest curve in the left of these plots is along a principal axis, and is clearly different from the others, with
$z\simeq z_{FRW}$.  The directions of the other trajectories have a
range of random angles and, although they start off with different
gradients, they very rapidly converge upon the mean.  As shown above,
deviations from the mean depend on the number of domains traversed. In Figure \ref{ZStat} we plot
the relative deviations from the mean, and find that for Milky Way sized masses
they do indeed become negligible very rapidly.  There is very little
scatter at $z \gtrsim 0.1$.

We can now use our numerical analysis to show the limitations of
the analytic approximations. These are shown for comparison in the right-hand panel of
Figure \ref{Znum1}.  Although not perfect, it is can be seen that
(\ref{ZZFRW}) with $\langle \gamma \rangle =2/3$ offers a reasonable approximation.  With
$\langle \gamma \rangle =7/10$, however, we find that Equation
(\ref{ZZFRW}) fits the numerical results almost perfectly.  We will
therefore use this semi-analytic value for $\langle \gamma \rangle$ in our calculations, later on.

\begin{figure}[t]
\center \epsfig{file=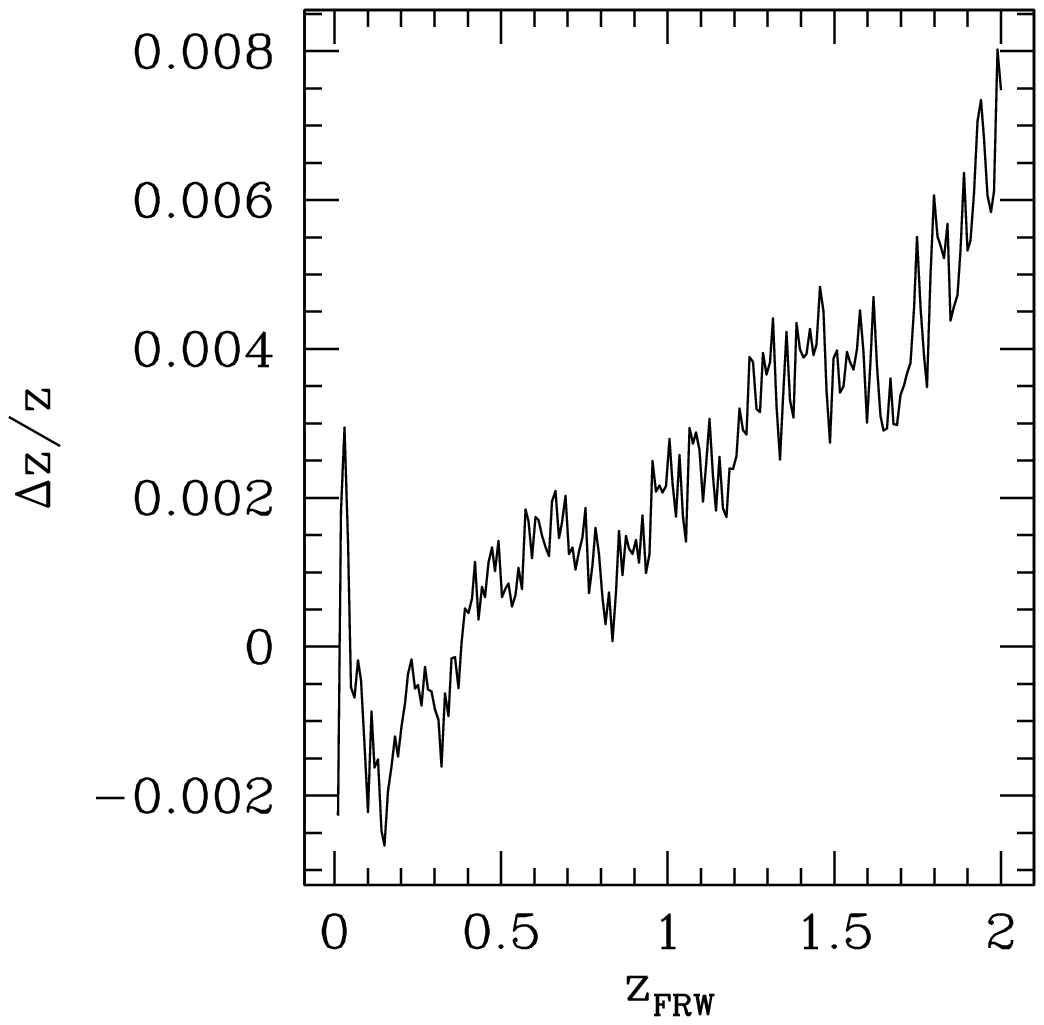,height=9cm} 
% \vskip -0.3in
\caption{{\protect{\textit{The relative difference in the $z-z_{FRW}$ relation for the two choices of matching
conditions at the cell boundaries.}}}}
\label{match}
\end{figure}

It is also now possible to compare the two matching schemes we
outlined above, for propagating geodesics between cells. As advertised,
we find that the effect on physical quantities is negligible, showing
that the approximate tangency of 3-spaces is indeed a valid concept. In
Figure \ref{match} we show this explicitly with a plot of $\Delta z/z$ as a function of
$z_{FRW}$, for the two different choices.  The
relative error is at the level of less than $1\%$ out as far as
$z_{FRW}\simeq 2$, which is remarkably accurate.

\begin{figure}[t]
\center \epsfig{file=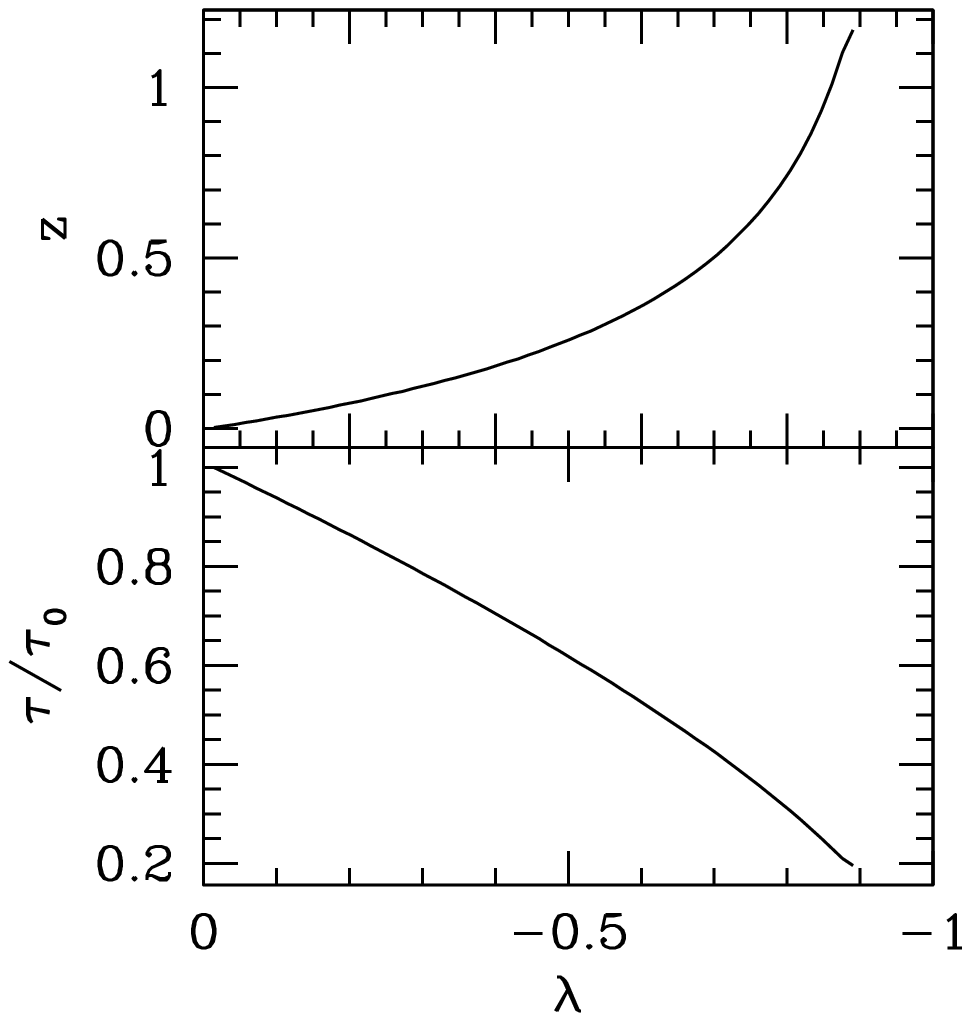,height=10cm} 
% \vskip -0.3in
 \caption{{\protect{\textit{The functional dependence for both $z=z(\lambda)$ and $\tau=\tau(\lambda)$, averaged
 over a set of $40$ runs.}}}}
\label{Figtau}
\end{figure}

In what follows it will also be useful to know the cosmological time,
 $\tau$, and redshift, $z$, as functions of the affine distance along
 the geodesics, $\lambda$. One finds, both analytically and numerically, that 
\begin{eqnarray}
\frac{\tau}{\tau_0}\simeq \left(1+\lambda \right)^{\frac{3}{2 \langle
    \gamma \rangle +3}}, \label{taulam}
\end{eqnarray}
where $\tau_0$ is the current age of the Universe and $\lambda$ has
been chosen so that it is $0$ at $\tau=\tau_0$, and $-1$ at $\tau=0$. In Figure \ref{Figtau} we plot
both $\tau=\tau(\lambda)$ and $z=z(\lambda)$ all the way from the
initial singularity ($\lambda=-1$) to the present day ($\lambda=0$).
These results are derived from averaging over a set of $40$
realisations.  We find that the numerical results fit Equation
(\ref{taulam}) almost perfectly.

\section{Optics of a Discretized Matter Distribution}
\label{5}

While the dynamical evolution of a cosmological model is important for
understanding, for example, the growth of structure, one could argue
that it is of equal, or even greater importance, to understand its
optical properties.  While the latter requires a knowledge of the
former, there is no guarantee that similar `average' dynamical properties in two
space-times should lead to similar optical properties.  In fact, we
have already seen that the lattice space-time gives different redshifts
to FRW, and, as we discussed in the introduction, there are good reason
to suspect that there should be considerable differences in distance measures too.

There are various reasons to suspect this.  Firstly, the dynamical equation for the expansion of a
bundle of null geodesics is driven by a term $\propto R$, the Ricci
curvature tensor.  In the continuous fluid approximation this is non-zero,
while in a universe with discrete islands of matter it will be
exactly zero everywhere outside of the matter itself. What is more, the shear in a homogeneous and isotropic
perfect fluid universe is zero, while in a universe with discrete
matter sources it will be non-zero.
These differences have competing effects on the luminosity
of astrophysical sources, the former making them dimmer in the lattice
universe, and the latter making them brighter.  These phenomena have been described by
Bertotti \cite{Bert}.

To investigate the optical properties of the space-time we will have
to integrate the Sachs optical equations \cite{Sachs} along each null trajectory.  These equations read
\bea
\frac{d \tilde{\theta}}{d \lambda} +\tilde{\theta}^2-\omega^2 + \sigma^*
\sigma &=& -\frac{1}{2}R_{ab} k^a k^b\\
\label{vort}
\frac{d \omega}{d \lambda} + 2 \omega \tilde{\theta} &=& 0\\
\frac{d \sigma}{d \lambda} +2 \sigma \tilde{\theta} &=& C_{a b c d}
(t^*)^a k^b (t^*)^c k^d,
\eea
where $\tilde{\theta}$, $\omega$ and $\sigma$ are the expansion,
rotation and complex shear scalars, respectively.  The $C_{a b c d}$
is Weyl's tensor, $R_{ab}$ is the Ricci tensor, and $t^a$ is a vector that is orthogonal to
$k^a$, null, and has a magnitude of $1$ (i.e. $t^a k_a=0$, $t^a t_a=0$ and $t^a (t^*)_a=1$).  In
Schwarzschild space-time $R_{ab}=0$, and the driving term in the $\sigma$ equation can
then be found to be
\be
\label{Cphase}
C=C_{a b c d} (t^*)^a k^b (t^*)^c k^d = \frac{3 m J^2}{r^5}  e^{i\Psi},
\ee
where $\Psi$ is a constant, specifying the complex phase.
Once the expansion scalar is known, then the angular diameter distance
is given by the integral
\be
\label{rA}
r_A \propto \text{exp}\left\{ \int_e^o \tilde{\theta} d \lambda \right\},
\ee
and the luminosity distance is given by Etherington's theorem
\cite{eth} as
\be
r_L = (1+z)^2 r_A.
\ee
It can also be seen that for a point source $\omega =0$ is always an
integral of (\ref{vort}). Choosing units appropriately we then have the two equations
\bea
\label{rA2}
\frac{1}{r_A} \frac{d^2 r_A}{d \lambda^2} + \sigma^*
\sigma &=& 0\\
\label{sig}
\frac{d \sigma}{d \lambda} + \frac{2 \sigma}{r_A} \frac{d r_A}{d
 \lambda} &=& \frac{3 m J^2}{r^5} e^{i\Psi} .
\eea
The initial conditions for integrating these equations are then
$\sigma \vert_o =0$, $r_A \vert_o =0$ and $d r_A/d\lambda \vert_o
=\text{constant}$.  (The reader should not confuse $r$ and $r_A$ in these
equations).  The value of $r=r(\lambda)$ should be substituted from
the solution to the geodesic equations, together with the relevant
value of $J$ and $\Psi$ in each cell.

Let us now write the complex shear as
\be
\sigma \equiv \vert \sigma \vert e^{i \Phi},
\ee
and define the new variable $X \equiv \vert \sigma \vert r_A^2$.  This
expression for $\sigma$ has two independent real parts, $\vert \sigma
\vert$ and $\Phi$, and therefore contains all the information about
the shear scalar.

%Now consider the driving term in the shear evolution equation,
%(\ref{sig}).  The phase, $\Psi$, of this term can be found from the
%complex vector $t^a$, as we will now show.  It will also be shown that $\Psi=\Psi_0=$constant, in any given
%cell\footnote{In the globally Schwarzschild case one can choose initial
%  conditions such that $\Psi_0=0$.  Here this can be done for one cell, but not all.}.

In terms of these new variables the evolution equations for the
optical scalars, (\ref{rA2}) and (\ref{sig}), become
\be
\label{rA4}
r_A^3 \frac{d^2 r_A}{d \lambda^2} + X^2=0,
\ee
and
\begin{align}
\label{dx}
\frac{dX}{d \lambda} &= \frac{3 m J^2 r_A^2}{r^5} \cos (\Psi-\Phi)\\
\label{dphi}
X\frac{d\Phi}{d\lambda} &= \frac{3 m J^2 r_A^2}{r^5} \sin (\Psi-\Phi),
\end{align}
for the real and imaginary parts of (\ref{sig}), respectively.  These
last two equations can then be integrated to give
\be
\label{Xexit}
\frac{X}{X_0} = \frac{\sin(\Psi_0 -\Phi_0)}{\sin(\Psi-\Phi)},
\ee
where $X_0$, $\Psi_0$ and $\Phi_0$ are constants. This results in the
single equation for the evolution of the shear
\be
\label{sig2}
\left( \frac{dX}{d\lambda} \right)^2 = \left( \frac{3 m J^2
 r_A^2}{r^5} \right)^2 \left( 1 -\frac{X_0^2 \sin^2 (\Psi_0-\Phi_0)}{X^2} \right),
\ee
where the sign of $dX/d\lambda$ is given by the sign of
$\cos(\Psi-\Phi)$.  Once we know $X_0$, $\Phi_0$ and $\Psi_0$ in a given cell, we can then
solve the equations (\ref{rA4}) and (\ref{sig2}), above.  

\subsection{Evolving shear between cells}

We now want to know how to evolve shear between cells.  It should
be the case that observable quantities are continuous along
trajectories, so we know that $r_A$ and $X$ should be the same on
starting the next cell as they were on leaving the last one.  It
remains to determine how the phase factors $\Phi$ and $\Psi$ are
propagated across cell boundaries.

It is clear from (\ref{rA4})-(\ref{sig2}) that the difference between
$\Phi$ and $\Psi$ is of primary importance for evolving the
optical scalars.  To make clear what this quantity corresponds to, let
us consider further the 4-vectors $t^a$.  We can think of them as
giving a set of two mutually orthogonal
space-like unit vectors, that are orthogonal to $k^a$, via
\be
t^a = \frac{1}{\sqrt{2}} (p^a +i q^a),
\ee
such that $p^ap_a=q^aq_a=1$ and $p^ak_a=q^ak_a=p^aq_a=0$.  The phases
$\Phi$ and $\Psi$ therefore correspond to the orientation of the shear
and of the driving term (\ref{Cphase}) in the plane spanned by $p^a$
and $q^a$.  The quantity $\Psi-\Phi$ is their relative orientation.

Now, in a space-time that is globally Schwarzschild (i.e. with only
one point-like mass in the whole space-time) it can be seen from
(\ref{dphi}) that $\Phi$ is driven toward $\Psi$.  This means that
the direction that the beam is sheared in is drawn into alignment with
the term driving the shearing, just as should be expected.  In the
present situation, however, when we pass between cells the orientation
of the shear that has accumulated up until that point, $\Phi$, will not, in
general, be in the same direction as the term driving the shear in the
new cell, $\Psi$.  Stated another way, the beam is sheared in different
directions as it passes through the different cells.  This change of
direction is given by the change in $\Psi-\Phi$ as the beam passes a cell
boundary.

We are, of course, free to perform an arbitrary change of $t^a$ as we move
between cells.  If we were to do this, however, then we should expect to have to work out
both $\Phi$ and $\Psi$ with respect to the new set of 4-vectors.
Instead, it makes sense to not rotate $p^a$ and $q^a$ at the boundary,
so that we can keep the phase of accumulated shear up until that
point, $\Phi$, the same.  This can be achieved by a direct transform
from the coordinate system of the first cell, $x^a$, into the
coordinate system of the second, $x^{\hat{a}}$, via $x^{\hat{a}} =
x^{\hat{a}} (x^a)$.  The vector $t^a$ then transforms as
\be
\label{that}
\hat{t}^{\hat{a}}\vert_{\text{entry to new cell}} = \frac{\partial
 x^{\hat{a}}}{\partial x^b} t^b \vert_{\text{exit from old cell}}
\equiv t^{\hat{a}}\vert_{\text{exit from old cell}} .
\ee
This is the same coordinate transformation that was used previously on
$n^a$, and hats mean quantities in the cell with the hatted coordinate
 system.  We now need to know the phase of the driving term in the
 second cell, $\hat{\Psi}$, with respect to these vectors.

To find this, consider that for the Schwarzschild geometry (\ref{pan})
and tangent vector (\ref{kn}) we can write $t^a$, for $E=1$, as
\be
\label{generalt}
t^a = e^{-i \Psi/2} (\bar{t}^a + \alpha k^a),
\ee
where
\be
\label{tbar}
\bar{t}^a = \frac{r \dot{\tau}}{\sqrt{2} J} \left( i n^r ; i
 \left\{ 1+\sqrt{\frac{2 m}{r}} n^r \right\} , -\sin \theta n^{\phi},
\frac{n^{\theta}}{\sin \theta} \right),
\ee
and where $\Psi$ and $\alpha$ are real and complex functions,
respectively\footnote{They represent the three required degrees of
 freedom from the eight components of $t^a$, and the five conditions
 it must satisfy.}.  To see that this $\Psi$ is the same as that in (\ref{Cphase}),
simply contract it with the Weyl tensor using the result\footnote{The
 symmetry of $C_{abcd}=-C_{bacd}=-C_{abdc}$ means that the term
 involving $\alpha$ in $t^a$ does not contribute to $C$.}
\be
\bar{C}=C_{a b c d} k^a (\bar{t}^*)^b k^c (\bar{t}^*)^d = \frac{3 m J^2}{r^5}.
\ee
We now need to know how $\Psi$ and $\alpha$ evolve inside each cell.
To find this let us impose that $t^a$ is
parallelly propagated along the null curve.  This then gives
\be
\frac{D t^a}{d \lambda} = k^b {t^a}_{;b} = 
-\frac{i t^{a}}{2} \frac{d \Psi}{d\lambda} +e^{-i \Psi/2} k^a \left( \frac{i
 B}{\sqrt{2} J} +\frac{d\alpha}{d\lambda}\right) =0,
\ee
which results in
\be
\label{psialpha}
\Psi=\Psi_0=\text{constant} \qquad \qquad \text{and} \qquad \qquad
\alpha=\alpha_0-\frac{i B}{\sqrt{2} J} \lambda,
\ee
where $\Psi_0$, Re$(\alpha_0)$ and Im$(\alpha_0)$ are constants.  
These can initially be set to zero in the first cell, without loss of
generality, but in subsequent cells are generally non-zero.

Now, the tangent vector in the new cell, $k^{\hat{a}}$, is already known
from the previous section, and $\hat{\bar{t}}^{\hat{a}}$ is the same
as in (\ref{tbar}), but with $n^a$ replaced by $n^{\hat{a}}$.  The values of
$\hat{\Psi}_0$, Re$(\hat{\alpha}_0)$ and Im$(\hat{\alpha}_0)$ can then
be found from (\ref{that})\footnote{Equation (\ref{that}) is 4 equations for
3 variables, so it must be the case that only three of these 4 are
independent.}. Although $\hat{\alpha}_0$ is not needed directly for integrating
(\ref{sig2}), it is needed to find $\hat{\Psi}_0$, and so should be
recorded.  Its value at exit from cell two is found from
(\ref{psialpha}).  This gives us the required value of $\Psi-\Phi$ when
starting the new cell.  

The prescription for calculating shear is then the following:

\vspace{10pt}

(1)  At the beginning of the cell, take $X_0$, $\Psi_0$, $\Phi_0$ and $\alpha_0$
    for that cell.  These will have been calculated at the end of the
    previous cell (see steps below).  If this is the first cell then take
    $X_0=\Psi_0=\Phi_0=\alpha_0=0$.

\vspace{10pt}

(2)  Substitute $X_0$, $\Psi_0$ and $\Phi_0$ into (\ref{sig2}), and
    integrate from the beginning of the cell until the trajectory
    hits another edge.

\vspace{10pt}

(3)  Use this $X=X(\lambda)$ to integrate (\ref{rA4}) along the same
    trajectory, to find $r_A$.

\vspace{10pt}

(4)  We now want to calculate $\hat{X}_0$ and $\hat{\Phi}_0$ for the next cell (hats
    indicate quantities in the next cell).  $\hat{X}_0$ is given by $\hat{X}_0=X_e$,
    where subscript $e$ means evaluated at the exit from the cell.  The
    value of $\hat{\Phi}_0$ is given by $$\hat{\Phi}_0=\Phi_e=\Psi_0- \arcsin \left(
    \frac{X_0}{X_e} \sin (\Psi_0-\Phi_0) \right).$$

\vspace{10pt}

(5)  It remains to find $\hat{\Psi}_0$ and $\hat{\alpha}_0$.  These
    will be found from Eq. (\ref{that}).  Now, $t^{\hat{a}}$ (at the
    exit from the first cell) is known from (\ref{generalt}),
    (\ref{tbar}) and (\ref{psialpha}).  In evaluating $t^{\hat{a}}$
    take $B$, $J$, $\Psi_0$, $\alpha_0$ and $k^a$ all from the first
    cell, and use $r$ and $\lambda$ as appropriate at the cell
    boundary.  This gives the RHS of (\ref{that}).  The LHS is given
    by $$\hat{t}^{\hat{a}} = e^{-i \hat{\Psi}_0/2}
    (\hat{\bar{t}}^{\hat{a}}+\hat{\alpha}\hat{k}^{\hat{a}}).$$  The
    $\hat{\bar{t}}^{\hat{a}}$ and $\hat{k}^{\hat{a}}$ in this
    expression are different to $\bar{t}^{a}$ and ${k}^{{a}}$, and are not just a
    coordinate transformation of them.  The tangent vector
    $\hat{k}^{{a}}$ is the tangent vector in the new cell (evaluated
    at the boundary still).  This is the new vector found in the
    previous section.  The vector $\hat{\bar{t}}^{\hat{a}}$ has
    the same functional form as (\ref{tbar}), but now with all hatted
    quantities, as found in the new cell (including $J$ from the new
    cell).  

\subsection{Solving the equations.}

The influence of shear complicates the solving of the optical
equations (\ref{rA4}) and (\ref{sig2}).  We can, however, make quick
progress if we are prepared to make approximations.  The system we are
describing involves the propagation of geodesics through multiple
lattice cells.  At low redshifts, and for trajectories that do not pass
close by to a central mass, we expect the effect of the shear to be
small, as the driving term $C \sim 1/r^5$.  In the absence of shear,
when $\sigma \sim 0$, we can write the solution for $r_A$ in (\ref{rA2}) as
\be
\label{rA3}
r_A \simeq c_1 + c_2 \lambda.
\ee
At larger redshifts, and for trajectories that pass close to a central
mass, however, the influence of shear will accumulate and become non-negligible.

In a classic paper, Press and Gunn \cite{PressGunn} addressed the impact of condensed
objects on the optical properties of the Universe. They modelled the evolution of the optical equations in a clumpy universe
as a Markov process, with $\sigma$ performing a random walk through
sporadic scattering events. It was found that, apart from some
extreme scattering events, the cumulative effect of the source terms
in the Sachs equations is negligible unless the correlation length of the objects is very large,
or the distances travelled were very great. In fact, the change in shear over a Hubble length $H_0^{-1}$
should be $\sigma\simeq H_0^{3/2}$ and hence $\sigma/H_0\ll 1$.
We find that the main features derived in \cite{PressGunn} are, essentially,
in agreement with the numerical results that come from integrating the
optical equations through a lattice universe.

\begin{figure}[t]
\center \epsfig{file=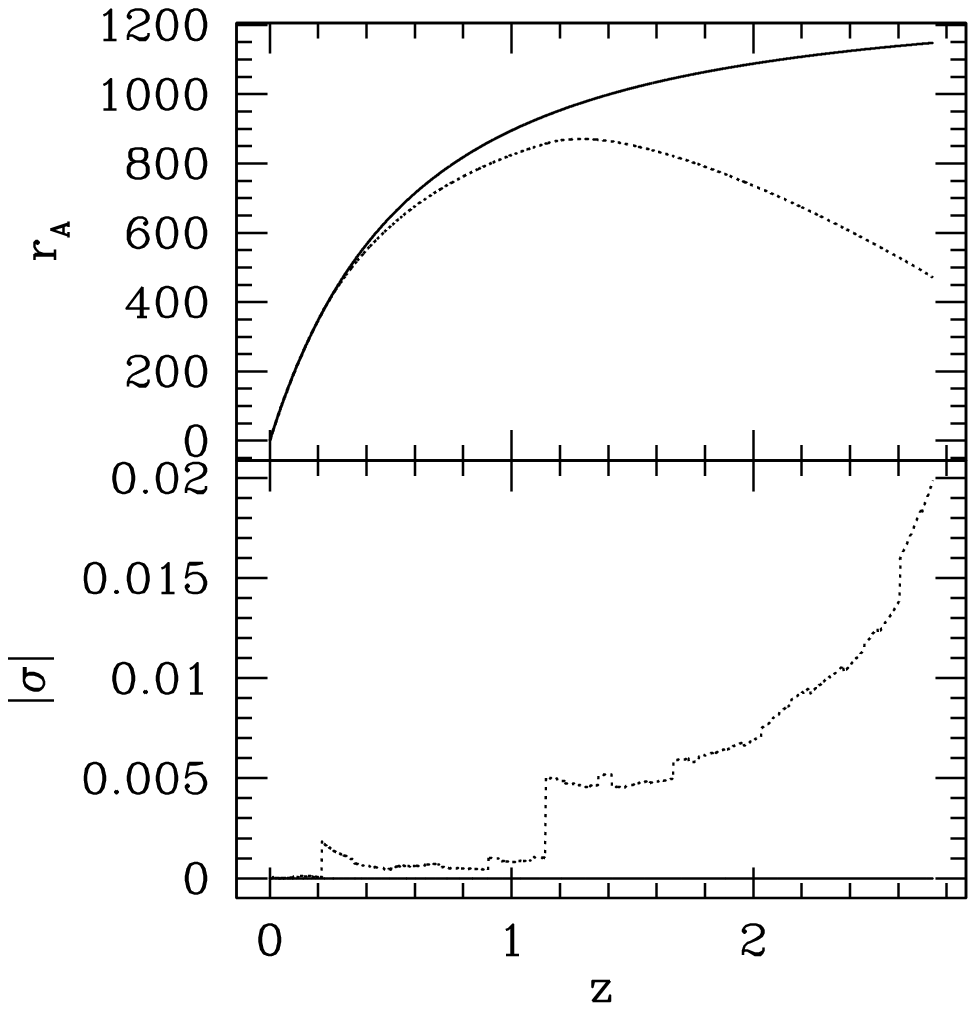,height=11.5cm} 
% \vskip -0.4in
 \caption{{\protect{\textit{The top panel shows the evolution of $r_A$ as a function of $z$ for a case
 where $\sigma$ is negligible $($top, solid line$)$ and for the case where it diverges $($bottom, dotted line$)$. 
 In the bottom panel we plot the evolution with redshift of the
 corresponding $\vert \sigma \vert$.
 }}}}
\label{Figshear}
\end{figure}

In Figure \ref{Figshear} we plot $r_A(z)$ for two different cases; one
in which $\sigma$ is always negligible,  and one in which $\sigma$ increases substantially. 
Of course, shear is always negligible initially.  Far from the origin the situation is more
complex, and $\vert \sigma \vert$ (or $X$) can deviate considerably
from zero. We find that 
$\vert \sigma \vert$ evolves as if it were subject to a Markov process, performing a random walk and slowly diverging
away from zero. Once it reaches a critical value, the angular diameter
distance, $r_A$, is found to stop growing, and to subsequently decrease, leading to
the formation of a caustic. The likelihood of this happening for a
random trajectory is small, but may become non-negligible
at moderate to high redshift, when many cells have been traversed.
The result, if such events do occur, is a substantial suppression of
$r_A$, corresponding to a significant brightening of distant sources
viewed along the trajectory in question.

We will now refine Press and Gunn's calculation, and work out how likely it will be for the shear to diverge and lead
to a turnover in $r_A$. To do so, we first note that there is a high
degree of regularity and recurrence and that, for the most part,
it should be acceptable to consider propagation through each domain as almost
independent from those that have come before. In each domain we rotate
the coordinate system so that the photon enters each cell in an
upward direction (i.e. through the $(x,y)$ plane). By approximating the lattice side as a disc
we can then write the point of entry into cell $i$ as 
\be
\nonumber
{\bf r}_i \equiv L{\bf
  q}_i=L(\hat{q}_{i}\cos\phi_i,\hat{q}_{i}\sin\phi_i,-1/2),
\ee
where $L$ is the side length of the cell, and the probability distributions for $\hat{q}_{i}$ and $\phi_i$
are
$P(\hat{q}_{i},\phi_i)d\hat{q}_{i}d\phi_i=\hat{q}_{i}d\hat{q}_{i}d\phi_i$,
where $\hat{q}_i\in[0,1/2]$ and $\phi_i\in[0,2\pi]$.
The direction vector at the point of entry can then be written
\be
\nonumber
{\dot {\bf r}}_i \simeq {\dot
  \tau}_i(\sin\chi\cos\xi,\sin\chi\sin\xi,\cos\chi),
\ee
where $\chi\in[0,\pi/2]$ and $\xi\in[0,2\pi]$.  The vector components
in the expressions above are given in a Cartesian coordinate system.

We now wish to find the approximate evolution of $X$. The $\dot{X}$
and $\dot{\Phi}$ equations tell us that $\Phi \rightarrow$constant, as
$X$ becomes large. We also know that $\Psi$ is constant in each
domain, but rotates randomly between domains.  We then have 
\be
X(\lambda)\simeq \sum_{i=0}^NA_i\cos(\Phi_i-\Psi_i) \nonumber
\ee
where
\be
A_i \equiv  3mJ^2_i\int_{\lambda_i}^{\lambda_{i+1}}d\lambda\frac{r_A^2}{r^5}. \nonumber
\ee
In this limit the trajectory $X=X(\lambda)$ can be seen to become
stochastic. 

Ultimately, we are interested in the case where a photon
crosses a large number of cells. The Central Limit Theorem
then implies that $X$ should have a Gaussian distribution. What we
want to know now is the mean and the variance of this distribution. To
find these consider that $\langle\cos(\Phi_i-\Psi_i)\rangle=0$ and
$\langle\cos(\Phi_i-\Psi_i)\cos(\Phi_j-\Psi_j)\rangle=\frac{1}{2}\delta_{ij}$.
We then have
\begin{eqnarray}
\langle X(\lambda)\rangle&=&0 \nonumber \\
\langle X^2(\lambda)\rangle&=&\frac{1}{2}\sum_{i=0}^{N} A^2_i\nonumber.
\end{eqnarray}
To proceed further we need to work out an approximate expression for $A_i$. If
the angular momentum in cell $i$ is given by $J_i^2=(\bf{r}_i \times
\bf{\dot{r}}_i )^2$ then we have 
\begin{eqnarray}
J_i^2=r^2_i{\dot r}^2_i-({\bf r}_i\cdot{\dot {\bf r}}_i)^2=L^2{\dot \tau}_i^2[q_i^2-({\bf q}_i\cdot{\bf n}_i)^2].
\end{eqnarray}
We now assume that we can approximate each geodesic as a straight
line. This gives
\begin{eqnarray}
{\bf r}&=&{\bf r}_i+{\dot {\bf r}}_i(\lambda_i-\lambda)={\bf r}_i-\lambda' {\dot {\bf r}}_i \nonumber \\
r^2&=&r_i^2+{\dot r}_i^2\lambda'^2-2{\bf r}\cdot{\dot {\bf r}}_i\lambda' \nonumber \\
r^2&=&L^2q_i^2+{\dot \tau}_i^2\lambda'^2-2{\dot \tau}_iL({\bf q}_i\cdot{\bf n}_i)\lambda'. \nonumber
\end{eqnarray}
We will assume that $\tau_i$ is approximately constant in each domain,
and set $r_A=r_{Ai}-{\alpha_i}\lambda'$. We then have that the
integral $A_i$ becomes 
\begin{eqnarray}
A_i=3mJ^2_i\int_0^{\Delta \lambda_i}\frac{r^2_A
  d\lambda'}{\left[L^2q^2+{\dot \tau}_i^2\lambda'^2-2{\dot
      \tau}_iL({\bf q}\cdot{\bf n})\lambda' \right]^{5/2}}. \nonumber 
\end{eqnarray}
If we change variable to $\Delta=-\alpha_i\lambda'$ then this becomes
\begin{eqnarray}
A_i=-\frac{3mJ^2_i}{\alpha_i}\int_0^{\Delta_i}\frac{(r_{Ai}+\Delta)^2d\Delta}{\left[L^2q^2+\left(\frac{{\dot
	\tau}_i}{\alpha_i}\right)^2\Delta^2+2\left(\frac{{\dot
	\tau}_i}{\alpha_i}\right) 
L({\bf q}\cdot{\bf n}) \Delta\right]^{5/2}}. \nonumber
\end{eqnarray}
The further change of variables to $Y=\frac{{\dot \tau}_i}{L\alpha_i}\Delta$ then gives
\begin{eqnarray}
A_i=-3mJ^2_i\frac{\alpha_i^2}{{\dot
    \tau}^3_i}\frac{1}{L^2}\int_0^{\Delta
  Y_i}\frac{(Y_{Ai}+Y)^2dY}{\left[(q^2+Y^2+2({\bf q}\cdot{\bf n})
    Y\right]^{5/2}}, \nonumber 
\end{eqnarray}
where we have assumed $L \simeq$constant over one domain, and defined $Y_{Ai}= \dot{\tau}_i r_{Ai}/\alpha_i L$.
The upper limit of integration is $\Delta Y_i=-\frac{{\dot \tau}_i}{L\alpha_i}{\alpha_i}\delta\lambda_i=
\frac{{\dot \tau}_i\Delta\lambda_i}{L}$, which we can approximate to 1. The integral is therefore time independent,
and depends only on the stochastic variables ${\bf q}$ and ${\bf n}$ (except for the presence of $Y_{Ai}$). Note, however,
that in the limit of multiple scattering events we have that $Y\ll Y_{Ai}$, so that 
\begin{eqnarray}
\int_0^{\Delta Y_i}\frac{(Y_{Ai}+Y)^2dY}{\left[(q^2+Y^2+2({\bf q}\cdot{\bf n}) Y\right]^{5/2}} \simeq 
Y_{Ai}^2\int_0^{\Delta Y_i}\frac{dY}{\left[(q^2+Y^2+2({\bf q}\cdot{\bf n}) Y\right]^{5/2}}. \nonumber
\end{eqnarray}
We then have that
\be
A_i = -\frac{3 m {\dot \tau}_i {\bar \alpha}^2 \lambda^2}{L^2}{\cal F}({\bf q},{\bf n}),
\ee
where
\begin{eqnarray}
&& {\cal F}({\bf q},{\bf n}) \nonumber \\ &=& [q^2-({\bf q}\cdot{\bf
      n})^2]\int_0^{1}\frac{dY}{\left[(q^2+Y^2+2({\bf q}\cdot{\bf n})
      Y\right]^{5/2}} \nonumber \\ 
&=& \frac{(q^2)^{\frac{3}{2}} (1+{\bf q} \cdot {\bf n} ) (2+3 q^2 +4
      {\bf q} \cdot {\bf n} -({\bf q} \cdot {\bf n} )^2)+ {\bf q}
      \cdot {\bf n}  (1+q^2+2 {\bf q} \cdot {\bf n} )^{\frac{3}{2}} (
      ({\bf q} \cdot {\bf n} )^2-3 q^2)}{3 (q^2)^{\frac{3}{2}}
      (1+q^2+2 {\bf q} \cdot {\bf n} )^{\frac{3}{2}} (q^2-({\bf q}
      \cdot {\bf n} )^2)}, \nonumber 
\end{eqnarray}
and we have taken $r_A= \bar{\alpha} \lambda$.
If we now take $\dot{\tau}_i=\dot{\tau}_0/a$ and $L=L_0 a$ then we can write the variance of $X$ as
\be
\langle X^2 \rangle = \frac{9 m^2 \dot{\tau}_0^2 \bar{\alpha}^4}{2 L_0^4} \sum_{i=0}^N \frac{\lambda^4 {\cal F}^2}{a^6}.
\ee
The continuum limit of this is
\be
\langle X^2 \rangle = \frac{9 m^2 \dot{\tau}_0^2 \bar{\alpha}^4
  \langle  {\cal F}^2\rangle}{2 L_0^5} \int_{\tau}^{\tau_0}
\frac{\lambda^4}{a^7} d\tau',
\ee
where we have used $\Delta i = - \Delta \tau_i/L$, and taken the
infinitesimal limit so that $di=-d\tau /L$.  Now with $\langle \gamma
\rangle = 7/10$ we have that $a=(\tau/\tau_0)^{2/3}$ and $\beta \lambda = ((\tau/\tau_0)^{22/15}-1)$,
which allows us to perform the integral above to find 
\begin{eqnarray}
{\cal G} &\equiv& \int_{\tau}^{\tau_0} \frac{\lambda^4}{a^7} d\tau' \nonumber \\
&=& \frac{\tau_0}{11 \beta^4} \left[ 3 \left(
  \frac{\tau_0}{\tau} \right)^{\frac{11}{3}} - 20 \left(
  \frac{\tau_0}{\tau} \right)^{\frac{11}{5}} +90 \left( \frac{\tau_0}{\tau}
  \right)^{\frac{11}{15}} + 60 \left( \frac{\tau}{\tau_0}
  \right)^{\frac{11}{15}} -5 \left( \frac{\tau}{\tau_0}
  \right)^{\frac{11}{5}} - 128 \right]\nonumber \\ 
&=& \frac{\tau_0}{11 \beta^4} \hat{{\cal G}}. \nonumber
\end{eqnarray}
We can then finally write
\be
\langle X^2 \rangle = \frac{9 m^2 \dot{\tau}_0^2 \bar{\alpha}^4 \langle  {\cal F}^2\rangle \tau_0}{22 L_0^5 \beta^4} \hat{{\cal G}}(\tau).
\ee

Let us now determine an estimate for the critical value of $X$ at
which the shear becomes important, and the divergence from the
background evolution occurs.  First we define the area of a bundle of
geodesics focused at the observer to be $Y \equiv r_A^2$.  This gives
the equation for $\ddot{r}_A$, (\ref{rA4}), as 
\be
\frac{Y \ddot{Y}}{2} = \frac{\dot{Y}^2}{4} - X^2 \nonumber .
\ee
Now let us take as our benchmark for the 
point at which divergence begins to be $\ddot{Y}=0$. Recall that $\ddot{Y} \simeq$constant
in the absence of shear, and goes from being initially positive (when
shear is negligible) to being negative when shear causes the turn over
and eventual divergence of $r_A$.  Taking $\ddot{Y}=0$ seems like as
reasonable place as any to mark the separation of these two regimes.
At this point we then find the critical value of $X$ to be
\be
X_c = \frac{\dot{Y}}{2} = r_A \dot{r}_A = \frac{r_A^2}{\lambda} =
\bar{\alpha}^2 \lambda = \frac{\bar{\alpha}^2}{\beta} \left( \left(
\frac{\tau}{\tau_0} \right)^{\frac{22}{15}}  -1   \right), 
\ee
where we have used $r_A =\bar{\alpha} \lambda$, as above (although this is only
strictly true when shear is completely negligible).  The ratio of $\langle X^2 \rangle$ to $X_c^2$
is then given by 
\be
\frac{\langle X^2 \rangle}{X_c^2} = \frac{9 m^2 \dot{\tau}_0^2 \langle
  {\cal F}^2\rangle \tau_0}{22 L_0^5 \beta^2} \frac{\hat{{\cal
      G}}}{\left( \left( {\tau}/{\tau_0} \right)^{\frac{22}{15}}  -1
  \right)^2}, 
\ee
where $ \langle  {\cal F}^2\rangle\simeq 10$ and $\hat{{\cal G}} (\tau)$ is
given by the expression above.  When this ratio becomes greater than
$1$, about $1/3$ of the trajectories will begin to
diverge\footnote{Under the assumption of the Central Limit Theorem.}.

\begin{figure}[t]
\center \epsfig{file=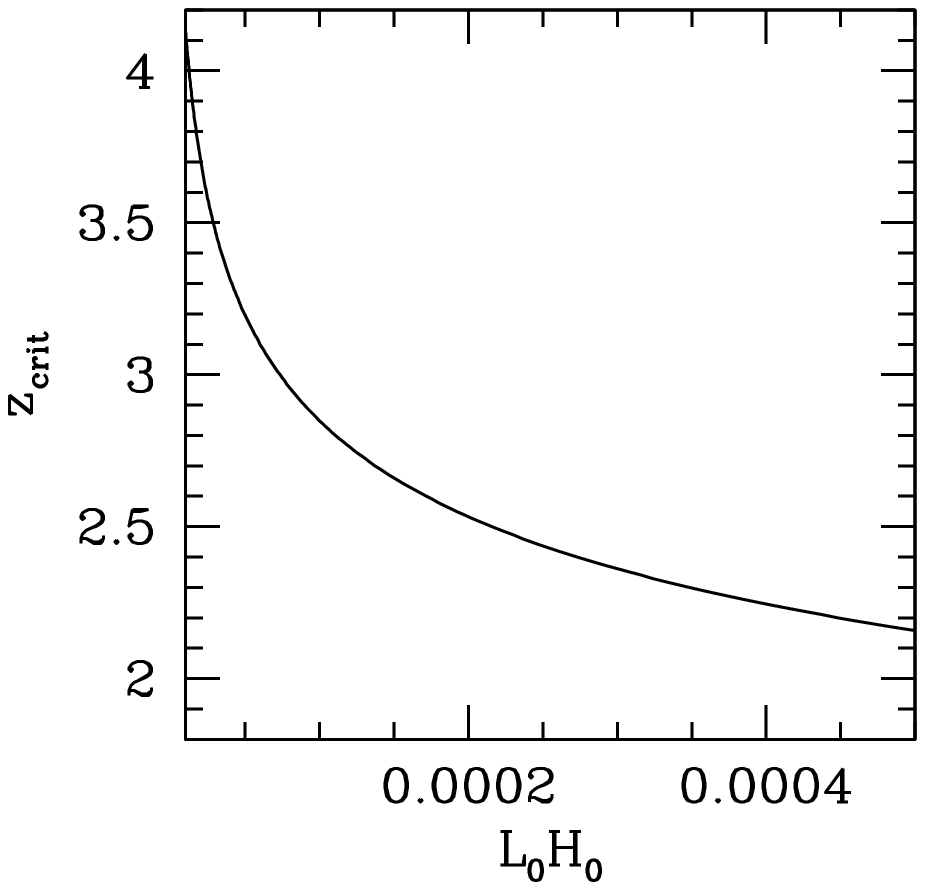,height=8cm} 
% \vskip -0.6in
 \caption{{\protect{\textit{The critical redshift, $z_{crit}$, as a
 function of the ratio of mass separation, $L_0$, to Hubble scale,
 $H_0$. At $z= z_{crit}$ we expect $\sim
 1/3$ of the trajectories will have experienced enough shear to cause a
 caustic.  For Milky Way size galaxies $L_0H_0\simeq 10^{-4}$.
 }}}}
\label{critredshift}
\end{figure}

In Figure \ref{critredshift} we plot the critical redshift as a
function of the ratio of mass separation, $L_0$, to Hubble scale,
$H_0$. We find that, in the regime that we are considering,
only a tiny fraction of trajectories should experience enough shear to cause
the optical scalars to diverge. For Milky Way sized galaxies we find
that the critical redshift is $z_{crit}\simeq  2.8$, and up until $z\simeq 1$
it should be an excellent approximation to take $\sigma\simeq 0$.

\section{Cosmological Observables}
\label{6}

Having discussed the dynamical and optical properties of an
Archipelagian Universe in some detail, let us now consider the
effects such a model of the Universe has on the interpretation of
cosmological observables.  Clearly, the optical properties of the
Universe are of great importance for very many different types of
cosmological observable.  These include reconstruction of the Hubble
diagram, interpretation of anisotropies in the CMB, and relating distances to redshifts for
the purpose of understanding galaxy number counts and baryon acoustic
oscillations.  In this section we will consider how these types of
observables will differ from FRW, in the model currently under
consideration.

\subsection{Distance measures, and the Hubble diagram}

In this section we will focus on the reconstruction of the Hubble
diagram in a universe with discrete matter content.
In the previous section we argued that shear will play a negligible
role in the optical equations, at least out to redshifts $\sim 1$.
Very few supernovae have been measured beyond this value, and 
hence we feel justified in using Equation (\ref{rA3}). 

If we enforce the condition that $\lambda=0$ at the observer then
$c_1=0$, and (\ref{rA3}) gives $r_A \propto \lambda$.
We  now want to relate $\lambda$ to $\tau$.  It was found above that
\be
1+z =\frac{\dot{\tau}_e}{\dot{\tau}_0} \simeq \left(
\frac{\tau_0}{\tau_e} \right)^{\frac{2 \langle \gamma \rangle}{3}}.
\ee
Integrating this gives
\be
\tau_e^{{\frac{2 \langle \gamma \rangle}{3}}+1}-\tau_0^{{\frac{2 \langle \gamma \rangle}{3}}+1} \propto \lambda,
\ee
where we have chosen the integration constant such that $\lambda
\rightarrow 0$ as $\tau_e \rightarrow \tau_0$.  This can now be
substituted into (\ref{t1}) to find $r_A \propto 1-(1+z)^{-1-3/2
  \langle \gamma \rangle}$. Etherington's theorem, $r_L = (1+z)^2 r_A$, then gives the luminosity
distance as a function of redshift as
\bea
r_L &=& (1+z)^2 -(1+z)^{ 1-\frac{3}{2 \langle \gamma \rangle} } \\
&\sim& z+ \left( 1- \frac{3}{4 \langle \gamma \rangle} \right) z^2 + \frac{(3 -2 \langle \gamma \rangle
)}{8\langle \gamma \rangle^2 } z^3
+ O(z^4),
\eea
where the constants of proportionality have been absorbed into $r_L$.
This can be compared to $r_L^{\text{dS}} = z+z^2$, $r_L^{\text{Milne}}
= z+z^2/2$ and $r_L^{\text{EdS}} = z+z^2/4-z^3/8+O(z^4)$.  

For $\langle \gamma \rangle=1$ $r_L$ is somewhere between Milne and EdS, and
is actually the same as was predicted by Bertotti \cite{Bert},
\be
\label{Bertrl}
r_L^{\text{Bertotti}} = z+\frac{(1-q_0)}{2} z^2 +\frac{q_0}{2} (q_0-f) z^3 + O(z^4),
\ee
with the deceleration parameter $q_0=1/2$ and the mass fraction\footnote{Taking
$f=1$ in (\ref{Bertrl}) gives the FRW distances supplied above.}
$f=0$.    In this case objects at the same $z$ appear dimmer than in
FRW.

\begin{figure}[t]
\center \epsfig{file=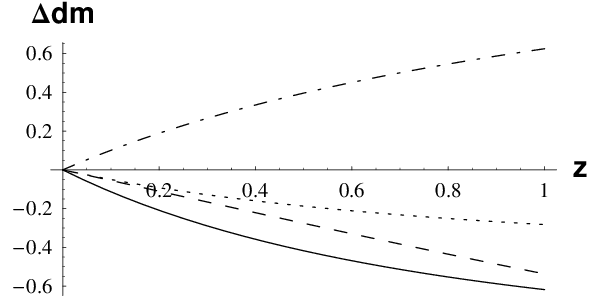,height=7cm} 
 \caption{{\protect{\textit{The distance modulus $\Delta dm$ for the
 Archipelagian Universe $($solid line$)$, an Einstein-de Sitter universe
 $($dashed line$)$, a de Sitter universe $($dot-dashed line$)$, and for a
 trajectory with $\langle \gamma \rangle =1$ $($dotted line$)$.
 }}}}
\label{distmod}
\end{figure}

For $\langle \gamma \rangle=2/3$ or $7/10$ the result is somewhat
different.  The lower redshift means that objects with the same $z$
now appear brighter than their FRW counterparts.  More generally, from
the above we can see that these models have an effective
deceleration parameter given by
\be
q^{lattice}_0= \frac{3}{2 \langle \gamma \rangle}-1,
\ee
in the absence of shear.
Using our semi-analytically determined value of $\langle \gamma
\rangle =7/10$ we find that
\bea
r_L&=& (1+z)^2 -(1+z)^{-\frac{8}{7}} \label{lumdistA}   \\
&\sim& z-\frac{1}{14}z^2 +\frac{20}{49}z^3 +O(z^4).
\eea
As advertised, the luminosity distance as a function of redshift is
such that the Universe is perceived to be decelerating at an even
more efficient rate than in FRW. Indeed, we
find $q_0=8/7$ for the Archipelagian Universe, as opposed to
$q_0=1/2$ for the Einstein-de Sitter model.  The luminosity distance
(\ref{lumdistA}) is shown graphically in Figure \ref{distmod}, in the
form of the distance modulus\footnote{Distance modulus is defined by
  $\Delta \text{dm} \equiv 5 \log_{10} (r_L/r_L^{\text{Milne}})$.}, $\Delta \text{dm}$.

\subsection{Other observables}

Hubble diagrams are, of course, not the only cosmological observables
that are sensitive to the optical properties of the Universe.  There
are various others, and indeed the CMB is famously an excellent probe
of the dynamics and optical properties of the Universe since the epoch
of last scattering, before which the Universe was opaque.  If we
assume the growth of structure in the Universe is well specified by
the usual perturbative analysis around an FRW background, then the
principle way in which the CMB will be sensitive to the new distance
relations found here will be in the relation between angles on the sky
today, and length scales at last scattering:  The projection of the
CMB onto our sky.

A useful probe of such effects on the CMB is the `shift parameter', which
specifies how much the peaks in the acoustic spectrum of primary
anisotropies are expected to move by.  The usual convention is to
specify this shift with respect to an EdS universe, in which case it
is simply given by the ratio of angular diameter distances in the test
cosmology and the corresponding distance in an EdS universe.  These
distances must be specified so that the Hubble rate at last scattering
is the same in each of the two universes, so that the physics up until
that point is also the same.

\begin{figure}[t]
\center \epsfig{file=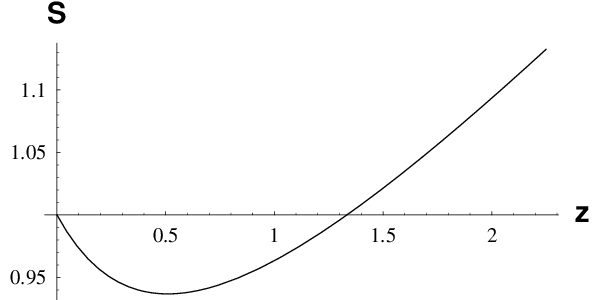,height=6cm} 
% \qquad \epsfig{file=S1.eps,height=5cm} 
\caption{{\protect{\textit{The shift parameter, from Equation
	$(\ref{Seq})$, as a function of $z$.  
}}}}
\label{Splot}
\end{figure}

From the considerations above we know that the angular diameter
distance in the Archipelagian Universe is, in the absence of shear, given by
\be
\label{Ara}
r_A^{lattice} = \frac{7}{22 H_0^{lattice}} \left[ 1-
  \frac{1}{(1+z)^{\frac{22}{7}}} \right].
\ee
The corresponding distance in EdS is given by
\be
\label{Seq}
r_A^{\text{EdS}} = \frac{2}{H_0^{\text{EdS}}} \left[ \frac{1}{(1+z)}-
  \frac{1}{(1+z)^{\frac{3}{2}}} \right].
\ee
As mentioned above, in this kind of procedure one would normally have to relate
$H_0^{\text{EdS}}$ to $H_0^{lattice}$, ensuring that in both space-times the
Hubble rate at last scattering was the same.  Here, however, we have that
the dynamics of the two space-times under consideration are, in fact,
the same.  We therefore have simply that $H_0^{\text{EdS}}= H_0^{lattice}$,
and so the shift parameter, $S$, is given by
\be
S \equiv \frac{r_A^{lattice}}{r_A^{\text{EdS}}} = \frac{7}{44}
\frac{(1-(1+z)^{-\frac{22}{7}})}{(1-(1+z)^{-\frac{1}{2}})} (1+z).
\ee
This is shown graphically in Figure \ref{Splot}.

It is clear that the CMB is very sensitive indeed to the type of
modifications to the usual distance relations that we are considering.
If we are conservative, and only apply (\ref{Ara}) out to a distance
of $z=2$, then we can see from Figure
\ref{Splot} that this already produces a shift in the CMB acoustic
spectrum of $\sim 10\%$.  This effect is already huge, with current
experiments able to constrain the shift to the order of $1\%$.  

Extrapolating our results to higher redshifts is somewhat tricky.
Of course, we know that at distances $z \gtrsim O(1)$ we should
include the effects of shear.  This will likely have a significant
effect on the angular diameter distance to last scattering.  Furthermore, the assumption
of Milky Way sized central masses for our cells will
likely become a poor approximation at high $z$, especially when we get
to the kind of redshifts before these structures even form.  As
mentioned previously, one may consider a different type of lattice in
which dark matter particles make up the central masses of much smaller
lattice cells, although the transition from galaxies to dark matter
particles will certainly be a highly non-trivial process.

For these reasons our numerical codes are currently unable to make predictions out to
the very high redshifts required to perform a proper analysis of the CMB, but
the discussion above indicates that there could be significant
deviations from the standard FRW picture.  We will leave further
considerations about CMB observations in these cosmologies, which will
likely be a highly non-trivial matter, to future publications.

As well as Hubble diagrams and the CMB there are a wealth of other
observables available to cosmologists.  Many of these, such as baryon
acoustic oscillations and galaxy number counts are highly sensitive
to the relation between cosmological time and redshift.  In our
models, this relation can be read off from Equation (\ref{ZZFRW}).
The smaller redshifts, that appear to be a generic prediction of this
model, mean that baryon acoustic oscillations should be measured at a
lower redshift in these models, as compared to the case of FRW.  In terms of
galaxy number counts, the lower redshift will mean that more spatial
volume will be included out to some particular $z$, and hence one
should expect a higher density of galaxies.

There are clearly many observational consequences of treating the
matter distribution as being discrete, rather than continuous.  It is
not the purpose of the current paper to work through all of these
rigorously, or to provide a complete working alternative to the
standard $\Lambda$CDM concordance model.  Rather, we have aimed at
gaining a decent understanding of the simplest dynamical and optical
properties of the simplest models.  Once again, more thorough
investigations of the
generalisations of these models, and their observational consequences,
will be postponed to future studies.

\section{Discussion}
\label{7}

In this paper we have considered an inhomogeneous cosmological model
with basic properties that are, in essence, similar to our own
Universe. The idea under-lying the model is that we should break away from
the doctrine that the Universe is permeated by a continuous fluid of
matter, and instead consider the view that it is predominantly empty space, punctuated by
islands of mass. We identify these islands as the {\it de facto} building
blocks of structure: Galaxies with a mass similar to that of the Milky
Way.

Our study extends and applies the ideas of the lattice universe
model proposed by Lindquist and Wheeler \cite{LW} that has, until now,
remained relatively unexplored.  We expect this approach to cosmology
to allow fresh insights into the effects of inhomogeneity in the
Universe, and, perhaps, to allow fresh approaches to understanding some of the
unresolved problems that have recently been revealed by observations.
The most notable of these is the evidence for Dark Energy.  Such
evidence relies heavily on understanding how to relate luminosities
and redshifts of distant astrophysical sources to the expansion
history of the Universe, and although we do not uncover any evidence that Dark Energy itself could
be explained by considerations of the type explored in this study,
this does not mean that refinements of the usual FRW cosmology will
not be of use for properly interpreting observational data\footnote{In
  \cite{next} we use Schwarzschild-de Sitter cells,
  instead of Schwarzschild cells, to construct a lattice universe.  In this way it is
be possible to study the effects of discretization of the matter content on estimates of
cosmological parameters in a more realistic fashion, by including the
effects of $\Lambda$.}.
Neither does it mean that future generalisations and refinements of this model
will not be more successful in this regard.  We consider the exploration of the optical properties of
different space-times to be essential to gaining a full understanding of
observations in cosmology.

Although we have chosen to study a highly symmetric model - evenly
spaced masses with a critical density - we believe that it does allow
us to reach some conclusions that should have consequences
for the real Universe.  For a start, the structure
of the Sachs optical equations are such that a bundle of null rays
travelling through empty space, and with negligible shear, should expand to give an angular
diameter distance $r_A \propto \lambda$.  This corresponds to sources at the
same affine distance being {\it dimmer} in the Archipelagian
Cosmology than they would be in the corresponding Einstein-de Sitter
universe.  It appears to us that this should be a robust generic prediction for {\it any}
space-time which is mostly empty space.  If the photons themselves do
not pass through the notional continuous fluid, then they do not
experience the extra focusing that such a fluid would produce.
Photons experience the integrated effects of the geometry through
which they pass, and {\it not} the `average' global geometry.

Similar effects have also been found in the context of weak
lensing. In most studies on this subject it is usual to take the
approach used by Press \& Gunn \cite{PressGunn} whereby one considers
a tube of FRW space-time, removes the evenly distributed mass, and
replaces it with the equivalent density of compact objects.  The
cosmological dynamics are then assume to remain FRW, while for the
purposes of integrating the optical equations one assumes a vacuum.  It
has been shown by Weinberg \cite{Weinberg} and Kibble \& Lieu
\cite{Kibble} that this approach leads to a distribution of magnitudes
with the same average as would have resulted from the continuous
energy density\footnote{See, however, \cite{EllisDunsby} who point out
the occurrence of caustics changes this result.}.  However, the distribution of magnitudes is
highly skewed \cite{Rauch}, with a large number of trajectories
experiencing little shear, and a small number experiencing large
shear due to the occurrence of rare lensing events. The sample mean of
a limited number of supernovae at small $z$ can therefore result in
deviations from the FRW average \cite{Marra}, and the skew can be used
as a test of the discreteness of the matter content \cite{Silk}.
Although our model is different to those just mentioned, our
results with regard to the effect of lensing do not appear to be in disagreement.

We also find, however, that redshifts can be altered from their FRW values.  We
have shown that, in the context of our model, the redshift can be
reasonably well understood in terms of the relative motion of the
boundaries of the lattice cells.  Unlike in an FRW cosmology, however,
it is only radial geodesics that experience redshifts that would
correspond to the Doppler shift due to the recessional
velocity of the global expansion.  All non-radial geodesics experience
smaller redshifts, as they see a locally anisotropic geometry, and
{\it not} the global `average'.  Despite the additional complications
that arise from this anisotropy, we find that after traversing a large
number of cells typical trajectories quickly approach a mean value $1+z\simeq
(1+z_{FRW})^{\frac{7}{10}}$.

These results allow us to then calculate observable measures of
distance as function of the redshift, as would be recorded by
astronomers.  We find that although objects appear dimmer at the same
affine distance, they appear {\it brighter} at the same redshift.
This is due to the redshift being lower to distant objects, as
discussed above.  Hence, we find that the distance moduli measured in a
critical density Archipelagian Universe should fare even worse 
when compared to supernova observations than a standard Einstein-de Sitter
universe. This is so even in the case of negligible shear, and can be
made worse when shear is included. Other cosmological observables are
also discussed, and the influence of the current considerations on
interpreting them is outlined.

An interesting effect that does not occur in FRW cosmology, but seems
to be an inevitable feature of Archipelagian Cosmology, is the
formation of caustics in the trajectories of photons.  The discrete
nature of the mass in this model means that shear, even if it is
negligible initially, will eventually accumulate, leading to the focusing, and
hence extreme brightening, of some distant sources. We find this effect to be
small out to redshifts of a few if the masses involved are sufficiently
small\footnote{For example, if they are Milky Way sized.}, but that it increases
substantial if we consider large agglomerations of mass, and large
redshifts.  Such considerations may therefore be important when
observing distant objects, and for CMB observations, in particular
(see \cite{EllisDunsby}).

The Lindquist-Wheeler model is highly idealised, and is only an
approximate solution of Einstein's equations. Nevertheless, it has
allowed us to draw conclusions about space-times which are qualitatively different from the standard
FRW space-time. An obvious next
step to improve the realism of the model is to consider irregular
lattices, such as those provided by, for example, a Voronoi tessellation. 
This is an ambitious proposal, and will undoubtedly throw up many difficulties
that the Lindquist-Wheeler model avoids.  However, it could allow us to
model the observed Universe in a new way, and provide a new
laboratory to study, for example, non-perturbative back-reaction.
Improving on the accuracy of the geometry as a solution of Einstein's
equations could be more difficult still, and the only way we can see to
make progress on this front is by using either numerically methods or weak
field approximations.

Finally, we should be careful to ensure that systematic errors have
not unduly influenced our results.  While it is routine to assume that
the vacuum space-time geometry outside mass concentrations
such as galaxies is well approximated by the Schwarzschild
solution, this is not exactly true: There will be external influences
from other gravitational sources that will deviate from perfect
spherical symmetry.  Determining the extent to which such influences
effect the result we have found here is likely to be a complicated matter, and
will be the subject of future publications.  One may also be concerned
as to the possibility of introducing systematic errors from the
matching conditions at cells boundaries.  Here we have used two different
methods to propagate trajectories between cells, and found that our
results are largely insensitive to which is chosen.  This suggests
that the boundary conditions are not a source of considerable error.
Furthermore, we have found analytic approximations that are in good
agreement with our numerical integrations, and that allow us
insight into the physical origin of the effects we have found.  We also show in a separate
publication that the familiar optical relations of de Sitter
space are recovered in the limit $\Omega_{\Lambda}\rightarrow 1$ \cite{next}.
We consider these results to suggest that the systematic
errors in our model are indeed under control. Confirming this by
determining the observational consequences of the approximations used
in this model will be the subject of future work.

In conclusion, we have shown that the differences between making
observations in a standard Einstein-de Sitter universe, and in a universe with discretized matter
content, can be substantial. A lesson we take away from this study is
that the assumption of continuity of energy in the Universe, while appearing innocuous, can affect our
interpretation of redshifts and luminosity distances considerably.
In order to make precise statements about the Universe, we should
therefore make sure we take these issues into account.

%\newpage

% ------------------------ ACKNOWLEDGEMENTS ----------------------------------
%\vspace{-10pt}
\section*{Acknowledgements}
%\vspace{-10pt}

We are extremely grateful to George Ellis for having introduced us to
the Lindquist-Wheeler model. We are also very grateful to Chris
Clarkson, Lance Miller, Francesco Sylos-Labini and Kane O'Donnell for helpful
discussions and comments.  TC is supported by Jesus College, and wishes to
acknowledge the BIPAC.

\appendix

\section*{A \quad Tiling the 3-space}

In order to consider a lattice model of the Universe, it is first necessary to
consider how to build a lattice out of tessellating cells that
fill the 3-space.  This problem is known as `tiling', and
has been considered in detail by Coxeter \cite{tiling}.

\begin{table}[htb]
\begin{center}
\begin{tabular}{|c|c|c|c|}
\hline
$\begin{array}{c} \bf{Lattice}\\
  \bf{Structure} \end{array}$ & $\begin{array}{c} \bf{Background}\\
  \bf{Curvature} \end{array}$ & $\begin{array}{c} \textbf{Cell}\\
  \textbf{Shape} \end{array}$
& $\begin{array}{c} \textbf{Number of}\\
  \textbf{Cells} \end{array}$ \\ 
\hline
\{333\} & + & Tetrahedron & 5 \\
\{433\} & + & Cube & 8 \\
\{334\} & + & Tetrahedron & 16 \\
\{343\} & + & Octahedron & 24 \\
\{533\} & + & Dodecahedron & 120 \\
\{335\} & + & Tetrahedron & 600 \\
\{434\} & 0 & Cube & $\infty$ \\
\{435\} & - & Cube & $\infty$ \\
\{534\} & - & Dodecahedron & $\infty$ \\
\{535\} & - & Dodecahedron & $\infty$ \\
\{353\} & - & Icosahedron & $\infty$\\
\hline
\end{tabular}
\end{center}
\caption{{\protect{\textit{Polyhedra that tile 3-surfaces of constant curvature, the
  number of cells required to fill the space, and the structure of
  the lattice $($in the form \{pqr\}$)$.  Hyper-spherical space is denoted by
  $`+$', flat 3-space by $`0$', and hyperbolic 3-space by $`-$'.  See the text for
  an explanation of \{pqr\}.  For further details see} \cite{tiling}.}}}
\label{table1}
\end{table}

In Table \ref{table1}, below, we show all the regular polyhedra that tile
all 3-spaces of constant curvature.  For the case of a positively
curved 3-space, as considered by LW, there are 6 possible tilings with
regular polyhedra.  These are with N=5, 8, 16, 24, 120 and 600 cells.
For flat 3-space there is only one possible tiling, consisting of an
infinite number of cubes.  Lastly, for negatively curved, hyperbolic
3-space, there are 4 polyhedra that can completely tile the space.
Lattices constructed on flat and negatively curved backgrounds have
been considered in a cosmological context by Redmount in \cite{red}.

The structure of a lattice can be well described using the compact
notation \{pqr\}, which is given for the possible tilings of 3-spaces
of constant curvature in Table \ref{table1}.  In this notation \emph{p}
denotes the number of edges on the face of a lattice cell, \emph{q}
denotes the number of faces that meet at the apex of any
individual cell, and \emph{r} denotes the number of cells that meet
at an edge.  Hence we have \{434\} for the cubic tiling of a flat space, as there are
$p=4$ edges to the square face of a cubic cell, $r=3$ squares meeting
at the corner of each individual cell (if it were considered in
isolation from the other cells), and $r=4$ cubes meeting around every
edge of every square.

For discussion of the efficacy of replacing the above polyhedra with spheres, the
reader is referred to \cite{LW}.

\section*{B \quad Geometric Set-up With a Cosmological Constant}
%\label{B}

It is straightforward to generalise the lattice cosmology to include a
non-zero $\Lambda$.  The metric in each lattice cell simply becomes
Schwarzschild-de Sitter, with the line-element
\be
\label{LS}
ds^2 = -\left(1-\frac{2 m}{r} -\frac{\Lambda}{3}r^2 \right) dt^2 + \frac{dr^2}{\left(
 1-\frac{2 m}{r} -\frac{\Lambda}{3}r^2 \right)} + r^2 d \Omega^2.
\ee
A coordinate system in which space-like surfaces, with $\tau=$constant,
are orthogonal to the world-lines of elements of radially free-falling
time-like shells is then given by the coordinate transformation (with $E>1$)
\be
\label{tau2}
dt = \frac{d \tau}{\sqrt{E}} + \frac{\left( E-\left(1- \frac{2 m}{r} -\frac{\Lambda}{3}r^2
 \right) \right) dr}{\left( 1- \frac{2 m}{r} - \frac{\Lambda}{3}r^2\right) \sqrt{E^2-E
   \left( 1-\frac{2 m}{r}-\frac{\Lambda}{3}r^2\right)}},
\ee
so that the line-element (\ref{LS}) becomes
\be
\label{pan2}
ds^2 = -\frac{1}{E}\left(1- \frac{2 m}{r}-\frac{\Lambda}{3}r^2 \right) d \tau^2 - \frac{2}{E}
\sqrt{E-\left(1-\frac{2 m}{r}-\frac{\Lambda}{3}r^2 \right)}
d\tau dr +\frac{dr^2}{E} +r^2 d\Omega^2,
\ee
%where $d\hat{s}= \sqrt{E} ds$.
Under the re-labelling $r\rightarrow a$, $m\rightarrow M/2$ and
$E\rightarrow 1-k$ we then have the equation of motion for a time-like
particle in radial free-fall being given by
\be
\frac{\dot{a}^2}{a^2} = \frac{M}{a^3}-\frac{k}{a^2}+\frac{\Lambda}{3},
\ee
which is the Friedmann equation for dust, with a cosmological
constant.

In this case the normalised four-velocity of a free-falling object is
\be
u^a= %\frac{1}{\sqrt{E}} 
\left( 1 ; \sqrt{E-\left( 1-\frac{2 m}{r}-\frac{\Lambda}{3}r^2
 \right) },0,0
 \right),
\ee
and it can be seen that such trajectories are orthogonal to surfaces
of constant $\tau$ as
\be
u^a n_a =0
\ee
for any arbitrary vector $n^a=(0;n^r,n^{\theta},n^{\phi})$ that exists
in such a surface.

\section*{C \quad Geodesic Equations in Cartesian Coordinates}
%\label{C}

For $E=1$ we can also express the geodesic equations in terms of Cartesian coordinates,
rather than spherical polars.  The line-element (\ref{pan}) then
appears as
\be
ds^2 = - \left( 1-\frac{2m}{r} \right) d\tau^2 - 2 \sqrt{\frac{2
    m}{r^3}} (x dx+y dy+z dz)d\tau + dx^2+dy^2+dz^2.
\ee
In terms of these variables the Euler-Lagrange equations read
\bea
\left( 1- \frac{2 m}{r} \right) \dot{\tau} + \sqrt{\frac{2 m}{r^3}} (x
\dot{x}+y \dot{y} +z\dot{z}) &=& B\\
\frac{\ddot{x}}{x}= \frac{\ddot{y}}{y}=\frac{\ddot{z}}{z} &=&
\sqrt{\frac{2 m}{r^3}} \ddot{\tau}-\frac{m}{r^3} \dot{\tau}^2,
\eea
with the null constraint
\be
\frac{2 m}{r^3} (x \dot{x}+y \dot{y} +z\dot{z})^2+\left(1-\frac{2
  m}{r} \right) (\dot{x}^2+\dot{y}^2+\dot{z}^2) = B^2.
\ee
These equations do not allow integrals as easily as in the spherical
polar case, but have fewer problems with coordinate singularities.

\section*{D \quad Geodesic Equations With a Cosmological Constant}
%\label{D}

The Euler-Lagrange equations for null geodesics in the space-time
specified by (\ref{pan2}) can be written
\begin{eqnarray}
\frac{d}{d\lambda} \left( \left(1-\frac{2 m}{r}-\frac{\Lambda}{3}r^2 \right) \dot{\tau} +
\sqrt{E-\left( 1-\frac{2 m}{r}-\frac{\Lambda}{3}r^2 \right)} \dot{r}
\right) &=& 0\\
\frac{d \dot{r}}{d\lambda}
- \sqrt{E-\left( 1-\frac{2
   m}{r}-\frac{\Lambda}{3}r^2\right)} \frac{d\dot{\tau}}{d\lambda} +
\left(\frac{m}{r^2} -\frac{\Lambda}{3}r \right)\dot{\tau}^2 &=&
E r \dot{\theta}^2+E r \sin^2 \theta \dot{\phi}^2 \\
\frac{d}{d\lambda} \left( r^2 \dot{\theta} \right) &=& r^2 \sin \theta \cos
\theta \dot{\phi}^2 \\
\frac{d}{d\lambda} \left( r^2 \sin^2 \theta \dot{\phi} \right) &=& 0,
\end{eqnarray}
together with the null constraint
\be
-\left(1-\frac{2 m}{r}-\frac{\Lambda}{3}r^2 \right) \dot{\tau}^2+ \dot{r}^2 +E r^2
\dot{\theta}^2+E r^2 \sin^2 \theta \dot{\phi}^2-2 \sqrt{E-\left(
 1-\frac{2 m}{r}-\frac{\Lambda}{3}r^2 \right)} \dot{r} \dot{\tau} =0.
\ee

\section*{E \quad Finding $\langle \gamma \rangle$}
%\label{E}

\begin{figure}[ht]
\center \epsfig{file=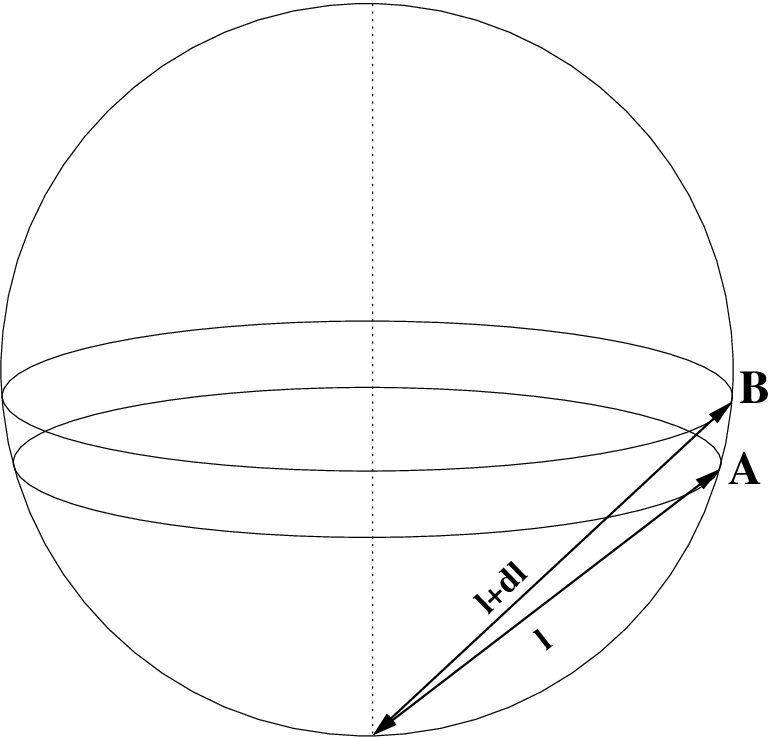,height=8cm}
\caption{{\protect{\textit{The spherical boundary of a cell.  We can consider the photon
to enter at the South pole, without loss of generality.}}}}
\label{fig1}
\end{figure}

\begin{figure}[htb]
\center \epsfig{file=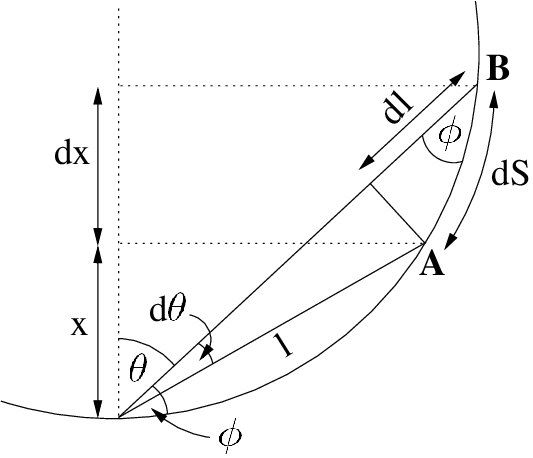,height=6.9cm} 
\caption{{\protect{\textit{A schematic of the chord joining the South pole and another
  random point on the surface of the sphere.}}}}
\label{fig2}
\end{figure}

To find $\langle \gamma \rangle$ consider a photon entering the spherical cell, rather than the
cube, as illustrated in Figures \ref{fig1} and \ref{fig2}.

Now consider a random set of trajectories in flat 3 dimensional space
(random refers to the measure of trajectories which is invariant under
translations, rotations and reflections).  These trajectories
represent the paths of photons in our model.  We now want to know the
distribution of chords that result from these trajectories
intersecting a unit sphere that is placed in this space.
It is shown in \cite{chord} that this situation is exactly equivalent
to selecting at random two points on the surface of the sphere and joining them with a
chord. In this case we can choose a coordinate system so that the first point is
at the South pole.  For the chord in question the ratio $\alpha/\beta$
is then given by $x/l=\cos \theta$, from Figure \ref{fig2}.
We will now proceed as prescribed by Berengut \cite{chord} to find the
mean and variance of $\cos \theta$.

The probability of finding our second point a chord length $l<L<l+dl$ away is given by the
area of the band in Figure \ref{fig1} over $4\pi$ (we consider
a unit sphere for now). To find the area, $A$, of this band consider
the cartoon schematic shown in Figure \ref{fig2}.  It can be seen that
\be
dl=dS \cos \phi = dS \sin \theta,
\ee
as $\phi=\pi/2-\theta$. The band area is then
\be
A = dS \times 2 \pi l \sin \theta  = 2 \pi l dl = 8 \pi \cos \theta
d\cos \theta,
\ee
as $l=2 \cos \theta$, and the probability of finding our second point in this band is
\be
P(\cos \theta<\cos \Theta<\cos \theta + d\cos \theta) = \frac{A}{4
  \pi}= 2 \cos \theta d\cos \theta,
\ee
giving the distribution $f(\cos \theta) = 2 \cos \theta$.  The mean and variance
of $\cos \theta$ are then given by $\overline{\cos \theta}=2/3$ and
$\sigma_{\cos \theta}^2=1/18$.  

The value of $\langle \gamma \rangle$ will then be given by the mean of
the sample of $n$ cells that the trajectory passes through, giving
\be
\label{beta}
\langle \gamma \rangle = \frac{2}{3}
\ee
with variance
\be
\sigma_{\langle \gamma \rangle}^2= \frac{1}{18 n}.
\ee
For a large number of cells this variance will soon become negligible, and
we will have $\langle \gamma \rangle \rightarrow 2/3$, as $n \rightarrow \infty$.

%\newpage

\end{document}